\documentclass[reprint,superscriptaddress,preprintnumbers,amsmath,amssymb,aps,prd,tightenlines,longbibliography]{revtex4-1}
\usepackage{graphicx}
\usepackage{xcolor}
\usepackage[utf8]{inputenc}
\usepackage{amsmath}
\usepackage{amssymb}
\usepackage{graphicx}
\usepackage{multirow}
\usepackage{comment}
\usepackage{slashed}
\usepackage{pifont}
\usepackage{xr-hyper}
\usepackage[colorlinks=true
,urlcolor=blue
,anchorcolor=blue
,citecolor=blue
,filecolor=blue
,linkcolor=red
,menucolor=blue
,linktocpage=true
,pdfproducer=medialab
,pdfa=true
]{hyperref}
\usepackage{units}

\newcommand{\med}[1]{\langle #1\rangle}

\newcommand{\mpl}{M_{\rm Pl}}

\def\L {\mathcal{L}}
\def \tr {{\rm Tr}}
\def \td {\theta}
\def \DM {\pi_{\rm DM}}

\begin{document}

\title{
  Dark matter in  QCD-like theories with a theta vacuum: cosmological and astrophysical implications
}

\author{Camilo Garc\'ia-Cely}
\affiliation{Instituto de F\'{i}sica Corpuscular (IFIC), Universitat de Val\`{e}ncia-CSIC, Parc Cient\'{i}fic UV, C/ Catedr\'{a}tico Jos\'{e} Beltr\'{a}n 2,
E-46980 Paterna, Spain}

\author{ Giacomo Landini}
\affiliation{Instituto de F\'{i}sica Corpuscular (IFIC), Universitat de Val\`{e}ncia-CSIC, Parc Cient\'{i}fic UV, C/ Catedr\'{a}tico Jos\'{e} Beltr\'{a}n 2,
E-46980 Paterna, Spain}

\author{ \'Oscar Zapata }
\affiliation{
Instituto de F\'isica, Universidad de Antioquia,\\
Calle 70 \# 52-21, Apartado A\'ereo 1226, Medell\'in, Colombia}

\begin{abstract}

QCD-like theories in which the dark matter (DM) of the Universe is hypothesized to be a thermal relic in the form of a dark pion has been extensively investigated, with most studies neglecting the CP-violating $\theta$-angle associated with the topological vacuum. 
 We point out that a non-vanishing $\theta$ could potentially trigger resonant number-changing processes giving rise to the observed relic density in agreement with perturbative unitarity as well as  observations of clusters of galaxies. This  constitutes a novel production mechanism of MeV DM and an alternative to those relying on the Wess-Zumino-Witten term.  Moreover, for specific meson mass spectra, similar resonant scatterings  serve as a  realization of velocity-dependent self-interacting DM  without a light mediator. Explicit benchmark models  are presented together with a discussion of possible signals, including gravitational waves from the chiral phase transition associated with the dark pions.

\end{abstract}

\maketitle

\begin{figure}[t]
\hspace{0.3cm}
\includegraphics[width =0.23\textwidth]{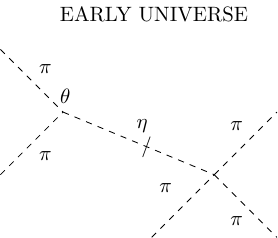} 
\\
\includegraphics[width =0.24\textwidth]{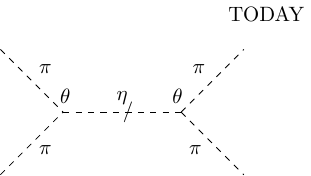}
\hspace{0.1cm}
\includegraphics[width =0.1\textwidth]{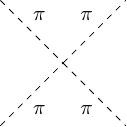}
\caption{
The $\theta$-vacuum could induce DM number changing processes in the early Universe and velocity-dependent self-scattering in DM halos today.
} 
\label{fig:fig1}
\end{figure}

Early studies of the strong-CP problem of the Standard Model (SM)~\cite{Shifman:1979if,Crewther:1979pi} found that the topological $\theta$-vacuum in QCD induces cubic interactions among the pseudo-Goldstone bosons of the meson octet, which are absent for $\theta=0$. In particular, the SM $\eta$ meson  may decay into two pions with a rate proportional to $\theta^2$.
Similar interactions for pion DM in QCD-like theories naturally induce number-changing processes in the Early Universe (see Fig.~\ref{fig:fig1}), which alter the DM amount that goes out of equilibrium from the primordial plasma. We will show that this constitutes a novel production mechanism of DM in QCD-like theories. This provides an alternative to the  popular SIMP model~\cite{Hochberg:2014kqa} of DM at the MeV scale based on $3\to2$  annihilations~\cite{Carlson:1992fn, Hochberg:2014dra}  induced by  a topological 5-point interaction, the so-called Wess-Zumino-Witten  (WZW) term~\cite{Wess:1971yu,Witten:1983tw}.  We refer the reader to~\cite{Hochberg:2015vrg, Kuflik:2015isi,Bernal:2015bla, Bernal:2015xba, Bernal:2015ova, Kuflik:2015isi, Choi:2015bya, Choi:2016hid,Soni:2016gzf, Kamada:2016ois,  Bernal:2017mqb,Cline:2017tka, Choi:2017mkk,  Kuflik:2017iqs, Heikinheimo:2018esa, Choi:2018iit,Hochberg:2018rjs, Bernal:2019uqr, Choi:2019zeb,  Katz:2020ywn, Smirnov:2020zwf, Xing:2021pkb, Braat:2023fhn, Bernreuther:2023kcg,Garani:2021zrr, Dey:2016qgf, Bernreuther:2023kcg}  for related work on MeV DM, and to ~\cite{Redi:2016kip, Draper:2018tmh, Abe:2024mwa} for studies of the effect of the $\theta$ angle in QCD-like DM models in other contexts.

Such $3\to2$ annihilations historically attracted significant attention because they predict DM collisions with a constant cross section per mass of order $\text{cm}^2/\text{g}$.  Known as self-interacting DM (SIDM)~\cite{Spergel:1999mh},  candidates with such cross sections could reconcile N-body simulations of collisionless cold DM --which predict universal halo profiles of large central densities~\cite{Dubinski:1991bm,Navarro:1995iw,Navarro:1996gj}-- with certain observations suggesting shallower profiles~\cite{Dave:2000ar,Vogelsberger:2012ku,Rocha:2012jg,Peter:2012jh,Elbert:2014bma,Fry:2015rta}. 
Nevertheless, over the years, it has become clear that SIDM models with a constant cross section do not fit the data well (see e.g.~\cite{Kaplinghat:2015aga,Tulin:2017ara}).  Concretely, it has been established that galaxy clusters constrain the cross section per unit mass  below 0.5~cm$^2$/g~\cite{Harvey:2015hha} (see also \cite{Sagunski:2020spe,  Harvey:2018uwf, Bondarenko:2017rfu, DES:2023bzs}), while larger values are required in small-scale objects where the velocity is smaller. 
In light of this, it is often believed that the only SIDM models that can explain the observed relic density are those involving a light mediator, which triggers the DM production and velocity-dependent self-interactions. See Refs.~\cite{Tulin:2017ara,Adhikari:2022sbh} for comprehensive reviews.

Meanwhile, the original SIMP model~\cite{Hochberg:2014kqa} is in tension with the cluster bound. Variants of this model have been proposed overcoming this tension (see, e.g. ~\cite{Kamada:2017tsq,Choi:2018iit,Bernreuther:2023kcg,Chu:2024rrv}), but without velocity-dependent SIDM. In this context, we find that the DM production mechanism induced by $\theta$ allows cross sections below the cluster bound. More importantly, we point out that for $\theta\neq0$, resonances similar to the $\eta$ meson mediate velocity-dependent scatterings, see Fig.~\ref{fig:fig1}. Hence, the $\theta$ angle may have astrophysical implications too.

We will illustrate this in two specific benchmark models, one of them resembling SM QCD.
We  structure the discussion as follows: the first section highlights the often-overlooked interactions induced by the  $\theta$ angle in QCD-like  models of DM, whereas the following two sections present ensuing cosmological and astrophysical consequences. Finally, we will present a summary along with a potential outlook. In the Appendix we provide the details of our
calculations.

\textbf{Pion DM in a $\theta$-vacuum. }
Although our conclusions apply to other gauge groups, we focus on  $SU(N_c)$ 
QCD-like theories of DM,  with $N_c\geq3$, and $N_f$ flavors of light Dirac quarks in the fundamental representation. As usual, the Lagrangian reads
\begin{equation}
	\L=-\frac{1}{4}F^2+\frac{ g^2\td}{32\pi^2}F\widetilde{F}+\bar{q}i\slashed{D} q-\left(\bar{q}_LM{q}_R+{\rm h.c.}\right).
 \label{eq:L}
\end{equation}

We assume $\td\ll1$ throughout for simplicity, and take
$M=\text{diag}(m_1,\cdots,m_{N_f})$, with  $m_q\neq0$ for all  flavors. Due to the chiral anomaly, under an arbitrary  transformation $q_{L,R}\to e^{\mp i\theta Q/2}q_{L,R}$, the angle in Eq.~\eqref{eq:L} shifts  as $\theta\to\theta(1-\tr\, Q)$ while 
$M\to M_{\td}= e^{i\td Q/2}M e^{i\td Q/2} $.
Taking transformations with $\tr\, Q=1$, we move the $\theta$ parameter  from the $F\widetilde{F}$ term to the quark mass matrix . 

In analogy with ordinary QCD, we expect that strong interactions confine at some  energy scale $\Lambda$,  
giving rise to a fermion condensate, $\med{\bar{q}q}\sim \Lambda^3$. This induces a chiral phase transition (PT) for  $m_q\ll\Lambda$,  
which spontaneously breaks the chiral flavor symmetry group of Eq.~\eqref{eq:L}:  $SU(N_f)_L\times SU(N_f)_R\to SU(N_f)_V$. This leads to $N_f^2-1$ pseudo-Goldstone bosons, $\pi^a$, described by  chiral perturbation theory (ChPT)~\cite{Pich:1991fq,Scherer:2002tk}, according to 
\begin{equation}\label{eq:chiralLag}
	\L_{\rm eff}=\frac{f_\pi^2}{4}\tr[\partial_\mu U^\dagger\partial^\mu U]+\frac{f_\pi^2}{2}B_0\tr[M^\dagger_\td U+ U^\dagger  M_\td],
\end{equation}
where $f_\pi$ is the dark meson constant, such that $\Lambda\sim  4\pi f_\pi/\sqrt{N_c}$, and
$U=\exp({i\pi^a\lambda^a/f_\pi})$
with $\lambda^a$ being the generators of $SU(N_f)$.
Here the condensate is parametrized as $\med{\bar{q}q} \equiv-B_0f_\pi^2$. 

The choice $Q=M^{-1}/\tr M^{-1}$ yields $\tr Q=1$ and no  linear terms in $\pi^a$  in the effective Lagrangian  at leading order in $\theta$~\cite{Crewther:1979pi,DiLuzio:2020wdo} 
\begin{equation}\label{eq:chiLagrangian}
\!\!\!\L_{\rm eff} \!\supset\frac{f_\pi^2B_0}{2} \left(\tr[M(U\!+\!U^\dagger)]\!-\!\frac{i\td}{\tr M^{-1}}\tr[U\!-\!U^\dagger]\right)\,.
\end{equation}
The first term includes  ordinary interactions involving an even number of mesons  such as	$\L_{\rm mass}=
 -B_0\pi^a\pi^b\tr[M\{\lambda^a,\lambda^b\}]/4$, to be corrected by ${\cal O} (\theta^2)$ contributions, while  
the second term --often neglected-- gives rise to interactions with an odd number of mesons\footnote{Throughout $\tr[\lambda_a\lambda_b]=2\delta_{ab}$, while the symmetric  tensors are  $ d_{abc}=\frac{1}{4}\tr(\{\lambda_a,\lambda_b\}\lambda_c)$ , and  $c_{abcde}=(\delta_{ab}d_{cde}+\delta_{cd}d_{abe})/N_f+\frac{1}{2}d_{abm}d_{cdn }d_{mne}$.}
\begin{equation}
\!\!\!\L_{\td}=-\frac{B_0\theta}{3f_\pi\tr M^{-1}}\bigg(d_{abc}\pi_a\pi_b\pi_c-\frac{c_{abcde}}{10f_\pi^2}\pi_a\pi_b\pi_c\pi_d\pi_e\bigg).
 \label{eq:thetaterms}
\end{equation}
Non-vanishing for $N_f\geq3$, this allows  processes absent when $\td=0$, particularly $s$-wave 3-to-2 annihilations, as pointed out in~\cite{Kamada:2017tsq} for degenerate meson spectra. This may be compared to the topological WZW term, 
$\mathcal{S}_{\mathrm{WZW}}= -i N_c\int \operatorname{Tr}\left(U^{\dagger} \mathrm{d} U\right)^5/240 \pi^2$,
which induces a 5-point interaction~\cite{Wess:1971yu,Witten:1983tw} leading to velocity suppressed  3-to-2 annihilations~\cite{Hochberg:2014kqa,Kamada:2022zwb}.  

For illustration, we consider the benchmark scenarios in table~\ref{table}. BM1 resembles ordinary QCD, with eight dark mesons referred to as in the SM. After accounting for the mixing angle, $\theta_{\eta\pi}$, the heaviest and lightest are  respectively $\eta$ and $\pi^0$, with $m_{\pi^0} \approx m_{\pi^\pm}< m_{K} < m_{\eta}$. 
On the other hand, 
BM2 has a remnant $SU(n)$ symmetry under which the $N_f^2-1$ mesons organize as  $n^2-1$ scalars in the adjoint, $2n$ scalars in the (anti)fundamental representations, $K/\bar{K}$,  and one singlet, the $\eta$ resonance. Neglecting $\mathcal{O}(\theta^2)$ corrections, the masses squared are respectively $2B_0 m ,\, B_0 (m+\mu)$ and $2B_0 \left(m+n \mu \right)/(n+1)$, with $\mu>m$.

\begin{table}[t]
\begin{tabular}{ccc}\hline
Benchmark &BM1 & BM2\\\hline
$N_f$ &3 &$n+1$ \\
$M$ & ${\rm diag}(m_u,m_d,m_s)$ & ${\rm diag}(m,..m,\mu)$ \\ 
Spectrum & $\pi^0,\pi^\pm,K^0,\Bar{K}^0,K^\pm,\eta$ & $\pi, K, \bar{K},\eta$\\
DM particle & $\pi^0\sim $ singlet &  $\pi \sim$ adjoint of $SU(n)$  \\
\hline
$\xi$ in Eq.~\eqref{eq:decay} &$\cos^2 3\theta_{\eta \pi}$&$6(n-1)/n $\\
$\med{\sigma_{\eta\pi} v}$ in Eq.~\eqref{eq:BEQfinal1} &$\frac{445\sqrt{5} m_{\rm DM}^2\delta}{5184\pi f_\pi^4}$& $\frac{\sqrt{5} m_{\rm DM}^2 (n^2-4)}{192 \pi  f_\pi^4 n^2 (n+1)}$
\\
$\sigma_0$  in Eq.~\eqref{eq:BW} &
$\frac{m_{\rm DM}^2}{128\pi f_\pi^4}$&
$\frac{m_{\rm DM}^2}{64\pi f_\pi^4}\frac{3n^4-2n^2+6}{n^2(n^2-1)}$
\\
\hline
\end{tabular}
\caption{Benchmark models considered in this work. 
}
\label{table}
\end{table}

\begin{figure*}[t]
\!\!\!\includegraphics[width=0.497\textwidth]{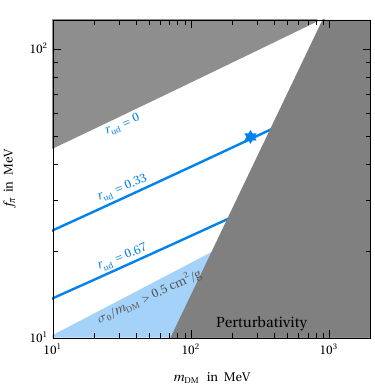}
\includegraphics[width=0.497\textwidth]{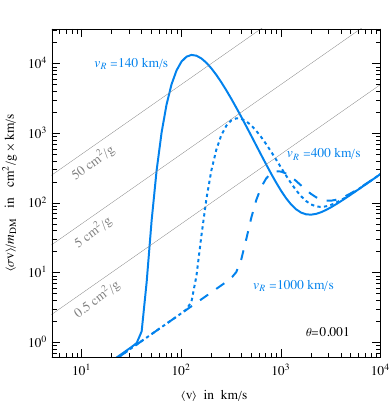} 
\caption{{\em Left:}  the cosmological DM relic abundance, $\Omega_{\rm DM}h^2=0.12$~\cite{Planck:2018vyg}, is reproduced along the plane  for different values of $r_{ud}$ in BM1. In the upper gray region DM is overabundant for any physical value of $r_{ud}$, while the blue region is excluded by observations of galaxy clusters, $\sigma_{0}/m_{\rm DM}>0.5\text{ cm}^2/\text{g}$. The  star corresponds to the benchmark points plotted on the right panel. In the dark-gray region ChPT breaks down, $m_{\rm DM}/f_\pi>4\pi/\sqrt{N_c}$~\cite{Kamada:2022zwb}. The inset shows the mass spectrum assuming $v_R,\, \theta \lesssim0.1$. {\em Right:}  Velocity dependence of the DM self-scattering cross section per unit of mass, for representative values of $v_R$ in Eq.~\eqref{eq:vR}.}  
 \label{fig:relic}
 \end{figure*}

Scaling  as $m_\pi\sim\sqrt{m_q\Lambda}$, meson masses are naturally close to each other  making plausible a spectrum with resonances. 
Dark baryons and dark glue-balls have typical masses  of order $\Lambda$. This justifies the assumptions that the DM is made of the lightest dark meson. All mesons are stable except $\eta$, as the cubic interaction in Eq.~\eqref{eq:thetaterms}  predicts  %
\begin{equation}\label{eq:decay}
\!\!\!\!\!\!\Gamma\left(\eta \rightarrow\! {\rm DM}\,{\rm DM}\right)
= \frac{\theta^2 B_0^2  \xi}{24 \pi f_\pi^2 m_\eta (\tr M^{-1})^2} \sqrt{1-\frac{4 m_{\rm DM}^2}{m_\eta^2}}\,,
\end{equation}
in agreement with~\cite{Shifman:1979if,Crewther:1979pi,Pich:1991fq}.  We introduce
\begin{equation}
    v_R = 2 \sqrt{\frac{m_\eta-2m_{\rm DM}}{m_{\rm DM}}} \,,
    \label{eq:vR}
\end{equation}
to study the non-relativistic resonant effects of Fig.~\ref{fig:fig1}, which appear for $v_R \lesssim 1$. For BM1 , up to $\mathcal{O}(v_R^2)$ and $\mathcal{O}(\theta^2)$ corrections, the relation between $ \delta\equiv m_{\pi^\pm}/m_{\pi^0}-1$ and $r_{ud}\equiv m_u/m_d$  is explicitly shown in the left panel of Fig.~\ref{fig:relic}, together with the corresponding mass spectrum.

\textbf{Cosmological implications.} The $\theta$ angle impacts DM production due to the number-changing processes resulting from Eq.~\eqref{eq:thetaterms}. These include 3-to-2 annihilations of DM and inverse decays ${\rm DM}\,{\rm DM} \to \eta$.  As discussed below, in the absence of the resonance, the former process in our benchmarks  only leads to the observed relic density in parameter space already excluded observationally. For the following discussion, we assume that the dark sector remains in kinetic equilibrium with the SM, thereby sharing the same temperature $T$.
The model-building requirements for this are similar to   those of ordinary SIMPs~\cite{Hochberg:2014kqa}. Interactions with the SM may allow (some of the) DM particles to decay. The requirements to avoid this are similar to those studied in the SIMP literature, see e.g.~\cite{Hochberg:2015vrg}. 

As in usual scenarios of co-annihilating DM, all the stable mesons eventually  convert to DM with a contribution to the DM relic density that depends on their mass. Accordingly, DM  and its coannihilating partners are referred to as $\DM$.  Kaons in both benchmarks are sufficiently heavy so that their contribution
is exponentially suppressed
and thus negligible. 
Consequently, $n_{\DM}\simeq n_\pi$ in BM2.
In contrast, due to $\delta \lesssim 0.075$, $\pi^{\pm}$ in BM1 give a non-negligible contribution   to the relic density, which can be calculated by noting that   
fast reactions $\pi^+\pi^-\leftrightarrow\pi^0\pi^0$ establish chemical equilibrium, $n_{\pi^0}/n_{\pi^\pm}=(n_{\pi^0}/n_{\pi^\pm})_{\rm eq}$, allowing to write a single Boltzmann equation for the combination $n_{\DM}=n_{\pi^0}+2n_{\pi^\pm}$. 
Note nevertheless that 
 $\pi^\pm$ convert into $\pi^0$ shortly after DM freeze-out. 
See~\cite{Katz:2020ywn} for an extended discussion. 

On the other hand, 
$\eta$ does not behave as a co-annihilating partner but as a catalyzer~\cite{Chu:2024rrv}, since  it induces number-changing processes
 such as $\eta\DM\to\DM\DM$, $\eta\to\DM\DM$   and their inverse. In light of this,
only Boltzmann equations for $\eta$ and $\DM$ are necessary.
 
They take a simple form in the case of most interest for this work, namely when $v_R \lesssim 0.1$ and $\theta$ is sufficiently large so that decays and inverse decays, $\eta\leftrightarrow\DM\DM$,  are both active even after all other DM number changing interactions have frozen-out. 
Then,
the process  $\DM\DM\DM\to \DM \DM$ is dominated by the exchange of an on-shell $\eta$ resonance (see Fig.~\ref{fig:fig1}). 
In practice, this means that the 3-to-2 annihilation is a two-step process: $\DM\DM \to \eta$ followed by $\eta\DM\to \DM\DM$, which justifies including only these in the Boltzmann equations and neglecting the small non-resonant piece~\cite{Kolb:1979qa}.

Noting that (inverse) decays do not affect $n=n_{\DM}+2n_\eta$, both equations combine giving $\dot{n}+3Hn=- \left( n_\eta n_{\DM}  \med{\sigma_{\eta\pi}v }- n_{\DM}^2   \med{\sigma_{\pi\pi}v} \right)$,
where  $\sigma_{\eta\pi} \equiv\sigma(\eta\DM\to \DM\DM)$, and similarly for the inverse process. 
This can be simplified  using detailed balance,  $ n^{\rm eq}_\eta\med{ \sigma_{\eta\pi} v} = n^{\rm eq}_{\DM}\med{\sigma_{\pi\pi}  v} $, as well as the fact that  
$\eta\leftrightarrow\DM\DM$ establish chemical equilibrium,  $ n_\eta/n_{\DM}^2=(n_\eta/n_{\DM}^2)_{\rm eq}$, or in terms of chemical potentials, $\mu_\eta=2\mu_{\DM}$.
 Putting everything together, the yield  for the combination $Y=Y_{\DM}+2Y_\eta$ satisfies
\begin{equation}\label{eq:BEQfinal1}
    \frac{dY}{dz}=-\med{\sigma_{\eta\pi} v}\frac{sY_{\eta,\rm eq}}{zH}\left(\frac{Y_{\DM}^3}{Y_{{\DM},{\rm eq}}^2}-\frac{Y_{\DM}^2}{Y_{{\DM},{\rm eq}}}\right),
\end{equation}
where as usual $Y_X=n_X/s$, $z\equiv m_{\rm DM}/T$, $H$
is the Hubble rate and $s$ the entropy density of the Universe.  The DM relic abundance is obtained upon numerical integration as $Y\simeq Y_{\DM} $. We show our results in the left panel of Fig~\ref{fig:relic} for BM1. These are independent of $v_R$ (if $v_R\lesssim0.1$),
but depend on 
$\delta$,  or equivalently on $r_{ud}$ if the small dependence of the masses on $v_R$ and $\theta$ is neglected. For BM2, depending on $N_f$, due to the large number of degenerate pions, the relic density is reproduced in parameter regions in tension with the cluster bound discussed below. 
This suggests that the DM  must lie in a small representation, which can be done breaking the mass degeneracy as in BM1. This is a generic feature of other color and flavor groups~\cite{future}. We also show the parameter region where ChPT holds, $m_{\rm DM}/f_\pi<4\pi/\sqrt{N_c} $.

Notice the relic density is independent of $\theta$ as long as
$\eta\leftrightarrow\DM\DM$ keep the two species in chemical equilibrium 
at least until the time at which $\eta\DM\to\DM\DM$  freezes out (see e.g.~\cite{Frumkin:2021zng}).
This implies a condition $\theta\geq\theta_{\rm min}(v_R,m_{\rm DM},f_\pi)$, estimated requiring that the thermally-average inverse decay rate 
overcomes the Hubble rate at freeze-out. 
We find $\theta_{\rm min}\sim 10^{-4}$ for both benchmarks. 
Notice also that the relic abundance is determined by $s$-wave 2-to-2 processes (see Table~\ref{table}), in contrast to the SIMP scenario based on velocity suppressed 3-to-2 annihilations.

When $\theta<\theta_{\min}$ or $v_R \gg 1$,  
the number changing interactions in Eq.~\eqref{eq:thetaterms} lead to 3-to-2 annihilation competing with the WZW term~\cite{Kamada:2017tsq}.
However, in our benchmark models,
these reproduce the observed DM relic density only in parameter regions excluded by cluster observations, to be discussed below.

\textbf{Astrophysical implications.} 
The $\theta$ angle could also induce resonant scattering among DM particles in present-epoch halos, see Fig.~\ref{fig:fig1}. The corresponding Breit-Wigner formula for the non-relativistic cross section is~\cite{Chu:2018fzy,Chu:2019awd}
\begin{eqnarray}
\label{eq:res}
\!\!\!\!\!\!\!\!\sigma(v)&=&\sigma_0+\frac{128\pi\,}{m_{\rm DM}^2 v_R^2 }\frac{\Gamma^2 }{ m_{\rm DM}^2\left( v^2 -  v_R^2 \right)^2+ 4\Gamma^2 v^2/v_R^2 }\,,
\label{eq:BW}
\end{eqnarray} 
which receives a resonant enhancement for DM relative velocities,  $v$, close to the mass combination, $v_R$, given in Eq.~\eqref{eq:vR}. For reference, in SI units, $v\sim$ 50 km/s in small-scale galaxies while $v\sim$2000 km/s in clusters of galaxies. The 1-loop $\eta$ self-energy  is included in $\sigma(v)$ as it impacts the scattering process~\cite{Chu:2018fzy}.   
$\sigma_0$ and other parameters for every benchmark are  reported in Table~\ref{table}.  We note that for BM1
all DM today consists of $\pi^0$ due to  the efficient conversion of $\pi^\pm$ into $\pi^0$ taking place shortly after freeze-out~\cite{Katz:2020ywn}.   Moreover, the production of $\pi^\pm$ from $\pi^0$ in astrophysical halos is kinematically forbidden  as  $\delta\gg v^2$ today  (which holds unless  $\delta\lesssim 10^{-5}$, or equivalently, $r_{ud}\sim1$).
For BM2, due to the $SU(n)$ symmetry,  $\sigma_0$ coincides with that reported in~\cite{Hochberg:2014kqa}.

N-body simulations of collisionless cold DM  predict universal halo profiles with large central densities~\cite{Dubinski:1991bm, Navarro:1995iw,Navarro:1996gj}, while certain observations allegedly suggest a shallower central DM density in small-scale halos. Accounting for DM scatterings in N-body  simulations indicates a possible solution, namely, a reduction of the central density in small-scale halos for  $\sigma(v)/m_{\rm DM}$ of several cm$^2/$g 
~\cite{Dave:2000ar,Vogelsberger:2012ku, Rocha:2012jg,Peter:2012jh,Elbert:2014bma,Fry:2015rta}.
This must be compared against the upper bound of $\sim 0.5$~cm$^2$/g from galaxy clusters.

In this context,  employing data from Ref.~\cite{Kaplinghat:2015aga}, Ref.~\cite{Chu:2018fzy} pointed out that the velocity dependence encoded in Eq.~\eqref{eq:BW} can successfully account for the required interactions in small-scale halos while evading the constraints arising from cluster observations. Following this study, in Fig.~\ref{fig:relic} (right panel)  we illustrate 
$\langle \sigma v\rangle/m_{\rm DM}$ for representative  parameter values 
yielding the observed DM density.   As is clear from the plot,  $v_R\sim100$ km/s gives a sharp velocity dependence, while  $v_R\sim1000$ km/s  gives practically no effect. We also verified that values of $\theta$ much larger than the one in Fig.~\ref{fig:relic} give a large self-interaction cross section  and are therefore disfavored~\footnote{Note that this justifies the expansion in $\theta$ of Eq.~\eqref{eq:chiLagrangian}.}. Nonetheless, we refrain from performing a global fit such as that presented in Ref.~\cite{Chu:2018fzy} because Ref.~\cite{Kaplinghat:2015aga} did not include aspects that are now known to play an important role but that are still under investigation (particularly for resonant SIDM~\cite{Tran:2024vxy,Kamada:2023dse}). This includes the effect of gravothermal collapse~\cite{Tran:2024vxy,Yang:2022hkm,Yang:2022zkd,Outmezguine:2022bhq} as well as observational data from ultra-faint dwarf galaxies~\cite{Hayashi:2020syu} or Milky-Way satellites~\cite{Valli:2017ktb, Correa_2021}. Nevertheless, we note that the behavior in Fig.~\ref{fig:relic} is favored by observations of ultra-faint dwarf galaxies, which suggest a sharp velocity dependence~\cite{Kamada:2023dse}.

Although in this letter we remain agnostic about the origin of the required values of $v_R$ by the SIDM hypothesis --of order $10^{-4}$ in natural units--
 we highlight that the $\theta$ angle itself could account for their smallness. For instance, any mechanism enforcing  $\mu=5m$ in BM2 leads to $v_R=0$ for $\theta=0$ and hence to $v_R \sim \theta$, as explained above. Furthermore, resonances with similar $v_R$ are found in nature, for example, in resonant collisions of $\alpha$-particles mediated by ${}^8\mathrm{Be}$. For other examples in QCD-like DM models, see~\cite{Tsai:2020vpi}. Also, while our DM production mechanism shares similarities with those induced by the $\rho$ resonance~\cite{Choi:2018iit} or DM bound states~\cite{Chu:2024rrv}, Refs.~\cite{Choi:2018iit,Chu:2024rrv} report no sizeable velocity-dependent self-interactions in halos.

A further signature is the emission of gravitational waves (GW) if the chiral PT is first order, see e.g.~\cite{Witten:1984rs,Schwaller:2015tja, Helmboldt:2019pan, Reichert:2021cvs}.
While in ordinary QCD the chiral PT is a smooth crossover~\cite{Aoki:2006we,Bhattacharya:2014ara}, studies of dark QCD sectors with three or more light fermions, $m_q\ll\Lambda$, establish that the PT may be first order~\cite{Schwaller:2015tja}, with a critical temperature $T_*\sim f_\pi$, potentially allowing for GW emission. For the specific case of $\theta=0$, utilizing several effective models, Ref.~\cite{Helmboldt:2019pan} presented QCD-like benchmarks for critical temperatures of order $T_*\sim 100$ MeV with a GW spectrum peaking at frequencies in the sub-mHz band (see also Ref.~\cite{Reichert:2021cvs}).
On the other hand, a nonvanishing $\theta$ may affect the PT: even with the matter content of SM QCD, $\theta\sim \pi$ could render the PT first order~\cite{Bai:2023cqj}, with potential consequences on the GW spectrum. This together with the fact that the dynamics of the PT and the GW spectrum depend on parameters and particle interactions not necessarily related with the pseudo-Goldstone bosons motivates a dedicated study for our benchmarks to draw conclusive statements. 
This is particularly important in light of the nHz signal from Pulsar Timing Arrays confirmed by the NANOGrav collaboration~\cite{EPTA:2023fyk,Reardon:2023gzh,NANOGrav:2023gor,Xu:2023wog}, which could be explained with a PT with $T_* \sim 10- 100$ MeV~\cite{NANOGrav:2023hvm}, in precise alignment with the viable values of $f_\pi$ in Fig. 2. See Ref.~\cite{Han:2023olf} for a hypothetical connection between the NANOGrav signal and SIDM with light mediators. The fact that our SIDM scenario does not require a light mediator could potentially alleviate the tension between this interpretation of the signal and cosmological observables requiring all unstable dark sectors particles to decay into the SM before $T\sim1$ MeV~\cite{Bringmann:2023opz}.

\textbf{Summary and outlook.}  We pointed out a production mechanism of DM in QCD-like theories where $\theta\neq0$. This requires a resonance mediating $\DM\DM \to \eta$ followed by $\eta\DM\to \DM\DM$, with masses such that  $v_R\lesssim 0.1$. Several resonances of this type are known to exist in the SM, particularly in the QCD sector~\cite{Tsai:2020vpi}.  Although it shares some similarities with theories where the relic density is obtained from annihilations induced by  WZW term, here the viable parameter regions can be reconciled with  cluster observations on DM self-scattering and are possibly  far from  the breakdown of ChPT,  see Fig.~\ref{fig:relic}. This mechanism works even for small values of the topological angle, $\theta\gtrsim10^{-4}$.

We also showed that the $\theta$-induced scatterings might address the apparent small-scale anomalies of the collisionless cold DM paradigm. As Fig.~\ref{fig:relic} shows, self-scattering cross sections per mass of several cm${}^2/$g are allowed in galactic halos, simultaneously with much smaller values in galaxy clusters. Thus, the $\theta$-vacuum leads to velocity-dependent SIDM without a light mediator,  provided that $v_R\sim$ 100 km/s. We emphasize that the relic density can be obtained with the mechanism mentioned above even if the SIDM hypothesis is not realized.

The $\theta$-dependent effects discussed here are generic features of QCD-like theories. For instance, in analogy with SM, Eq.~\eqref{eq:L} exhibits an anomalous axial $U(1)$ symmetry, whose pseudo-Goldstone boson, $\eta'$, has $\theta$-dependent cubic vertices~\cite{Witten:1980sp,DiVecchia:1980yfw}, potentially allowing for resonant scattering. 
Beyond $SU(N_c)$ gauge groups,  qualitatively similar effects arise for $SO(N_c)$ or $Sp(N_c)$.
In particular, with $N_c\geq4$, gauge confinement breaks flavor symmetry as $SU(N_f)\to SO(N_f)$ for $SO(N_c)$ or $SU(N_f)\to Sp(N_f)$ for $Sp(N_c)$. This leads to Lagrangian interactions analogous to Eq.~\eqref{eq:thetaterms} if $N_f\geq3$ for $SO(N_c)$ or $N_f\geq6$ for $Sp(N_c)$ (see the Appendix and~\cite{Sannino:2016sfx,LANDINIGIACOMO2022DMag, Buttazzo:2019mvl,Witten:1982fp}).
Models of this kind will be presented in an upcoming publication~\cite{future}, together with an analysis of the Boltzmann equations.
Additionally, to be explored in future investigations are the CP violation effects induced by the $\theta$-angle. Following Refs.~\cite{Cline:2017qpe, Carena:2018cjh} it would be possible to leverage the novel source of CP violation in QCD-like sectors with a $\theta$-angle to generate a matter-antimatter asymmetry. This approach would offer a potential workaround to the tight constraints imposed by searches for electric dipole moments in scenarios with CP-violation in the visible sector.

{\bf Acknowledgements.} We thank Xiaoyong Chu, Pilar Hern\'andez, Manoj Kaplinghat,  Hyungjin Kim and Tomer Volansky for useful discussions and Ayuki Kamada for comments on the manuscript.  C.G.C. is supported by a Ramón y Cajal contract with Ref.~RYC2020-029248-I, the Spanish National Grant PID2022-137268NA-C55 and Generalitat Valenciana through the grant CIPROM/22/69. G.L. is supported by the Generalitat Valenciana APOSTD/2023 Grant No. CIAPOS/2022/193. G.L. also acknowledges the hospitality of Universidade de Coimbra, Portugal. O.Z. has been partially supported by Sostenibilidad-UdeA, the UdeA/CODI Grants 2022-52380 and  2023-59130, and Ministerio de Ciencias Grant CD 82315 CT ICETEX 2021-1080.  O.Z. also acknowledges the support of the Simons Foundation under Award number 1023171-RC and the hospitality of the International Institute of Physics at the Universidade Federal do Rio Grande do Norte in Natal, Brazil.

\bibliographystyle{utphys-modified}
\bibliography{ref}

\providecommand{\href}[2]{#2}\begingroup\raggedright\begin{thebibliography}{10}

\bibitem{Shifman:1979if}
M.~A. Shifman, A.~I. Vainshtein, and V.~I. Zakharov, ``{Can Confinement Ensure Natural CP Invariance of Strong Interactions?},'' \href{http://dx.doi.org/10.1016/0550-3213(80)90209-6}{{\em Nucl. Phys. B} {\bfseries 166} (1980) 493--506}.

\bibitem{Crewther:1979pi}
R.~J. Crewther, P.~Di~Vecchia, G.~Veneziano, and E.~Witten, ``{Chiral Estimate of the Electric Dipole Moment of the Neutron in Quantum Chromodynamics},'' \href{http://dx.doi.org/10.1016/0370-2693(79)90128-X}{{\em Phys. Lett. B} {\bfseries 88} (1979) 123}. [Erratum: Phys.Lett.B 91, 487 (1980)].

\bibitem{Hochberg:2014kqa}
Y.~Hochberg, E.~Kuflik, H.~Murayama, T.~Volansky, and J.~G. Wacker, ``{Model for Thermal Relic Dark Matter of Strongly Interacting Massive Particles},'' \href{http://dx.doi.org/10.1103/PhysRevLett.115.021301}{{\em Phys. Rev. Lett.} {\bfseries 115} no.~2, (2015) 021301}, \href{http://arxiv.org/abs/1411.3727}{{\ttfamily arXiv:1411.3727 [hep-ph]}}.

\bibitem{Carlson:1992fn}
E.~D. Carlson, M.~E. Machacek, and L.~J. Hall, ``{Self-interacting dark matter},'' \href{http://dx.doi.org/10.1086/171833}{{\em Astrophys. J.} {\bfseries 398} (1992) 43--52}.

\bibitem{Hochberg:2014dra}
Y.~Hochberg, E.~Kuflik, T.~Volansky, and J.~G. Wacker, ``{Mechanism for Thermal Relic Dark Matter of Strongly Interacting Massive Particles},'' \href{http://dx.doi.org/10.1103/PhysRevLett.113.171301}{{\em Phys. Rev. Lett.} {\bfseries 113} (2014) 171301}, \href{http://arxiv.org/abs/1402.5143}{{\ttfamily arXiv:1402.5143 [hep-ph]}}.

\bibitem{Wess:1971yu}
J.~Wess and B.~Zumino, ``{Consequences of anomalous Ward identities},'' \href{http://dx.doi.org/10.1016/0370-2693(71)90582-X}{{\em Phys. Lett. B} {\bfseries 37} (1971) 95--97}.

\bibitem{Witten:1983tw}
E.~Witten, ``{Global Aspects of Current Algebra},'' \href{http://dx.doi.org/10.1016/0550-3213(83)90063-9}{{\em Nucl. Phys. B} {\bfseries 223} (1983) 422--432}.

\bibitem{Hochberg:2015vrg}
Y.~Hochberg, E.~Kuflik, and H.~Murayama, ``{SIMP Spectroscopy},'' \href{http://dx.doi.org/10.1007/JHEP05(2016)090}{{\em JHEP} {\bfseries 05} (2016) 090}, \href{http://arxiv.org/abs/1512.07917}{{\ttfamily arXiv:1512.07917 [hep-ph]}}.

\bibitem{Kuflik:2015isi}
E.~Kuflik, M.~Perelstein, N.~R.-L. Lorier, and Y.-D. Tsai, ``{Elastically Decoupling Dark Matter},'' \href{http://dx.doi.org/10.1103/PhysRevLett.116.221302}{{\em Phys. Rev. Lett.} {\bfseries 116} no.~22, (2016) 221302}, \href{http://arxiv.org/abs/1512.04545}{{\ttfamily arXiv:1512.04545 [hep-ph]}}.

\bibitem{Bernal:2015bla}
N.~Bernal, C.~Garcia-Cely, and R.~Rosenfeld, ``{WIMP and SIMP Dark Matter from the Spontaneous Breaking of a Global Group},'' \href{http://dx.doi.org/10.1088/1475-7516/2015/04/012}{{\em JCAP} {\bfseries 04} (2015) 012}, \href{http://arxiv.org/abs/1501.01973}{{\ttfamily arXiv:1501.01973 [hep-ph]}}.

\bibitem{Bernal:2015xba}
N.~Bernal and X.~Chu, ``{$\mathbb {Z}_2$ SIMP Dark Matter},'' \href{http://dx.doi.org/10.1088/1475-7516/2016/01/006}{{\em JCAP} {\bfseries 01} (2016) 006}, \href{http://arxiv.org/abs/1510.08527}{{\ttfamily arXiv:1510.08527 [hep-ph]}}.

\bibitem{Bernal:2015ova}
N.~Bernal, X.~Chu, C.~Garcia-Cely, T.~Hambye, and B.~Zaldivar, ``{Production Regimes for Self-Interacting Dark Matter},'' \href{http://dx.doi.org/10.1088/1475-7516/2016/03/018}{{\em JCAP} {\bfseries 03} (2016) 018}, \href{http://arxiv.org/abs/1510.08063}{{\ttfamily arXiv:1510.08063 [hep-ph]}}.

\bibitem{Choi:2015bya}
S.-M. Choi and H.~M. Lee, ``{SIMP dark matter with gauged Z$_{3}$ symmetry},'' \href{http://dx.doi.org/10.1007/JHEP09(2015)063}{{\em JHEP} {\bfseries 09} (2015) 063}, \href{http://arxiv.org/abs/1505.00960}{{\ttfamily arXiv:1505.00960 [hep-ph]}}.

\bibitem{Choi:2016hid}
S.-M. Choi and H.~M. Lee, ``{Resonant SIMP dark matter},'' \href{http://dx.doi.org/10.1016/j.physletb.2016.04.055}{{\em Phys. Lett. B} {\bfseries 758} (2016) 47--53}, \href{http://arxiv.org/abs/1601.03566}{{\ttfamily arXiv:1601.03566 [hep-ph]}}.

\bibitem{Soni:2016gzf}
A.~Soni and Y.~Zhang, ``{Hidden SU(N) Glueball Dark Matter},'' \href{http://dx.doi.org/10.1103/PhysRevD.93.115025}{{\em Phys. Rev. D} {\bfseries 93} no.~11, (2016) 115025}, \href{http://arxiv.org/abs/1602.00714}{{\ttfamily arXiv:1602.00714 [hep-ph]}}.

\bibitem{Kamada:2016ois}
A.~Kamada, M.~Yamada, T.~T. Yanagida, and K.~Yonekura, ``{SIMP from a strong U(1) gauge theory with a monopole condensation},'' \href{http://dx.doi.org/10.1103/PhysRevD.94.055035}{{\em Phys. Rev. D} {\bfseries 94} no.~5, (2016) 055035}, \href{http://arxiv.org/abs/1606.01628}{{\ttfamily arXiv:1606.01628 [hep-ph]}}.

\bibitem{Bernal:2017mqb}
N.~Bernal, X.~Chu, and J.~Pradler, ``{Simply split strongly interacting massive particles},'' \href{http://dx.doi.org/10.1103/PhysRevD.95.115023}{{\em Phys. Rev. D} {\bfseries 95} no.~11, (2017) 115023}, \href{http://arxiv.org/abs/1702.04906}{{\ttfamily arXiv:1702.04906 [hep-ph]}}.

\bibitem{Cline:2017tka}
J.~M. Cline, H.~Liu, T.~Slatyer, and W.~Xue, ``{Enabling Forbidden Dark Matter},'' \href{http://dx.doi.org/10.1103/PhysRevD.96.083521}{{\em Phys. Rev. D} {\bfseries 96} no.~8, (2017) 083521}, \href{http://arxiv.org/abs/1702.07716}{{\ttfamily arXiv:1702.07716 [hep-ph]}}.

\bibitem{Choi:2017mkk}
S.-M. Choi, H.~M. Lee, and M.-S. Seo, ``{Cosmic abundances of SIMP dark matter},'' \href{http://dx.doi.org/10.1007/JHEP04(2017)154}{{\em JHEP} {\bfseries 04} (2017) 154}, \href{http://arxiv.org/abs/1702.07860}{{\ttfamily arXiv:1702.07860 [hep-ph]}}.

\bibitem{Kuflik:2017iqs}
E.~Kuflik, M.~Perelstein, N.~R.-L. Lorier, and Y.-D. Tsai, ``{Phenomenology of ELDER Dark Matter},'' \href{http://dx.doi.org/10.1007/JHEP08(2017)078}{{\em JHEP} {\bfseries 08} (2017) 078}, \href{http://arxiv.org/abs/1706.05381}{{\ttfamily arXiv:1706.05381 [hep-ph]}}.

\bibitem{Heikinheimo:2018esa}
M.~Heikinheimo, K.~Tuominen, and K.~Lang\ae{}ble, ``{Hidden strongly interacting massive particles},'' \href{http://dx.doi.org/10.1103/PhysRevD.97.095040}{{\em Phys. Rev. D} {\bfseries 97} no.~9, (2018) 095040}, \href{http://arxiv.org/abs/1803.07518}{{\ttfamily arXiv:1803.07518 [hep-ph]}}.

\bibitem{Choi:2018iit}
S.-M. Choi, H.~M. Lee, P.~Ko, and A.~Natale, ``{Resolving phenomenological problems with strongly-interacting-massive-particle models with dark vector resonances},'' \href{http://dx.doi.org/10.1103/PhysRevD.98.015034}{{\em Phys. Rev. D} {\bfseries 98} no.~1, (2018) 015034}, \href{http://arxiv.org/abs/1801.07726}{{\ttfamily arXiv:1801.07726 [hep-ph]}}.

\bibitem{Hochberg:2018rjs}
Y.~Hochberg, E.~Kuflik, R.~Mcgehee, H.~Murayama, and K.~Schutz, ``{Strongly interacting massive particles through the axion portal},'' \href{http://dx.doi.org/10.1103/PhysRevD.98.115031}{{\em Phys. Rev. D} {\bfseries 98} no.~11, (2018) 115031}, \href{http://arxiv.org/abs/1806.10139}{{\ttfamily arXiv:1806.10139 [hep-ph]}}.

\bibitem{Bernal:2019uqr}
N.~Bernal, X.~Chu, S.~Kulkarni, and J.~Pradler, ``{Self-interacting dark matter without prejudice},'' \href{http://dx.doi.org/10.1103/PhysRevD.101.055044}{{\em Phys. Rev. D} {\bfseries 101} no.~5, (2020) 055044}, \href{http://arxiv.org/abs/1912.06681}{{\ttfamily arXiv:1912.06681 [hep-ph]}}.

\bibitem{Choi:2019zeb}
S.-M. Choi, H.~M. Lee, Y.~Mambrini, and M.~Pierre, ``{Vector SIMP dark matter with approximate custodial symmetry},'' \href{http://dx.doi.org/10.1007/JHEP07(2019)049}{{\em JHEP} {\bfseries 07} (2019) 049}, \href{http://arxiv.org/abs/1904.04109}{{\ttfamily arXiv:1904.04109 [hep-ph]}}.

\bibitem{Katz:2020ywn}
A.~Katz, E.~Salvioni, and B.~Shakya, ``{Split SIMPs with Decays},'' \href{http://dx.doi.org/10.1007/JHEP10(2020)049}{{\em JHEP} {\bfseries 10} (2020) 049}, \href{http://arxiv.org/abs/2006.15148}{{\ttfamily arXiv:2006.15148 [hep-ph]}}.

\bibitem{Smirnov:2020zwf}
J.~Smirnov and J.~F. Beacom, ``{New Freezeout Mechanism for Strongly Interacting Dark Matter},'' \href{http://dx.doi.org/10.1103/PhysRevLett.125.131301}{{\em Phys. Rev. Lett.} {\bfseries 125} no.~13, (2020) 131301}, \href{http://arxiv.org/abs/2002.04038}{{\ttfamily arXiv:2002.04038 [hep-ph]}}.

\bibitem{Xing:2021pkb}
C.-Y. Xing and S.-H. Zhu, ``{Dark Matter Freeze-Out via Catalyzed Annihilation},'' \href{http://dx.doi.org/10.1103/PhysRevLett.127.061101}{{\em Phys. Rev. Lett.} {\bfseries 127} no.~6, (2021) 061101}, \href{http://arxiv.org/abs/2102.02447}{{\ttfamily arXiv:2102.02447 [hep-ph]}}.

\bibitem{Braat:2023fhn}
P.~Braat and M.~Postma, ``{SIMPly add a dark photon},'' \href{http://dx.doi.org/10.1007/JHEP03(2023)216}{{\em JHEP} {\bfseries 03} (2023) 216}, \href{http://arxiv.org/abs/2301.04513}{{\ttfamily arXiv:2301.04513 [hep-ph]}}.

\bibitem{Bernreuther:2023kcg}
E.~Bernreuther, N.~Hemme, F.~Kahlhoefer, and S.~Kulkarni, ``{Dark matter relic density in strongly interacting dark sectors with light vector mesons},'' \href{http://arxiv.org/abs/2311.17157}{{\ttfamily arXiv:2311.17157 [hep-ph]}}.

\bibitem{Garani:2021zrr}
R.~Garani, M.~Redi, and A.~Tesi, ``{Dark QCD matters},'' \href{http://dx.doi.org/10.1007/JHEP12(2021)139}{{\em JHEP} {\bfseries 12} (2021) 139}, \href{http://arxiv.org/abs/2105.03429}{{\ttfamily arXiv:2105.03429 [hep-ph]}}.

\bibitem{Dey:2016qgf}
U.~K. Dey, T.~N. Maity, and T.~S. Ray, ``{Light Dark Matter through Assisted Annihilation},'' \href{http://dx.doi.org/10.1088/1475-7516/2017/03/045}{{\em JCAP} {\bfseries 03} (2017) 045}, \href{http://arxiv.org/abs/1612.09074}{{\ttfamily arXiv:1612.09074 [hep-ph]}}.

\bibitem{Redi:2016kip}
M.~Redi, A.~Strumia, A.~Tesi, and E.~Vigiani, ``{Di-photon resonance and Dark Matter as heavy pions},'' \href{http://dx.doi.org/10.1007/JHEP05(2016)078}{{\em JHEP} {\bfseries 05} (2016) 078}, \href{http://arxiv.org/abs/1602.07297}{{\ttfamily arXiv:1602.07297 [hep-ph]}}.

\bibitem{Draper:2018tmh}
P.~Draper, J.~Kozaczuk, and J.-H. Yu, ``{Theta in new QCD-like sectors},'' \href{http://dx.doi.org/10.1103/PhysRevD.98.015028}{{\em Phys. Rev. D} {\bfseries 98} no.~1, (2018) 015028}, \href{http://arxiv.org/abs/1803.00015}{{\ttfamily arXiv:1803.00015 [hep-ph]}}.

\bibitem{Abe:2024mwa}
T.~Abe, R.~Sato, and T.~Yamanaka, ``{Composite Dark Matter with Forbidden Annihilation},'' \href{http://arxiv.org/abs/2404.03963}{{\ttfamily arXiv:2404.03963 [hep-ph]}}.

\bibitem{Spergel:1999mh}
D.~N. Spergel and P.~J. Steinhardt, ``{Observational evidence for selfinteracting cold dark matter},'' \href{http://dx.doi.org/10.1103/PhysRevLett.84.3760}{{\em Phys. Rev. Lett.} {\bfseries 84} (2000) 3760--3763}, \href{http://arxiv.org/abs/astro-ph/9909386}{{\ttfamily arXiv:astro-ph/9909386}}.

\bibitem{Dubinski:1991bm}
J.~Dubinski and R.~G. Carlberg, ``{The Structure of cold dark matter halos},'' \href{http://dx.doi.org/10.1086/170451}{{\em Astrophys. J.} {\bfseries 378} (1991) 496}.

\bibitem{Navarro:1995iw}
J.~F. Navarro, C.~S. Frenk, and S.~D.~M. White, ``{The Structure of cold dark matter halos},'' \href{http://dx.doi.org/10.1086/177173}{{\em Astrophys. J.} {\bfseries 462} (1996) 563--575}, \href{http://arxiv.org/abs/astro-ph/9508025}{{\ttfamily arXiv:astro-ph/9508025}}.

\bibitem{Navarro:1996gj}
J.~F. Navarro, C.~S. Frenk, and S.~D.~M. White, ``{A Universal density profile from hierarchical clustering},'' \href{http://dx.doi.org/10.1086/304888}{{\em Astrophys. J.} {\bfseries 490} (1997) 493--508}, \href{http://arxiv.org/abs/astro-ph/9611107}{{\ttfamily arXiv:astro-ph/9611107}}.

\bibitem{Dave:2000ar}
R.~Dave, D.~N. Spergel, P.~J. Steinhardt, and B.~D. Wandelt, ``{Halo properties in cosmological simulations of selfinteracting cold dark matter},'' \href{http://dx.doi.org/10.1086/318417}{{\em Astrophys. J.} {\bfseries 547} (2001) 574--589}, \href{http://arxiv.org/abs/astro-ph/0006218}{{\ttfamily arXiv:astro-ph/0006218}}.

\bibitem{Vogelsberger:2012ku}
M.~Vogelsberger, J.~Zavala, and A.~Loeb, ``{Subhaloes in Self-Interacting Galactic Dark Matter Haloes},'' \href{http://dx.doi.org/10.1111/j.1365-2966.2012.21182.x}{{\em Mon. Not. Roy. Astron. Soc.} {\bfseries 423} (2012) 3740}, \href{http://arxiv.org/abs/1201.5892}{{\ttfamily arXiv:1201.5892 [astro-ph.CO]}}.

\bibitem{Rocha:2012jg}
M.~Rocha, A.~H.~G. Peter, J.~S. Bullock, M.~Kaplinghat, S.~Garrison-Kimmel, J.~Onorbe, and L.~A. Moustakas, ``{Cosmological Simulations with Self-Interacting Dark Matter I: Constant Density Cores and Substructure},'' \href{http://dx.doi.org/10.1093/mnras/sts514}{{\em Mon. Not. Roy. Astron. Soc.} {\bfseries 430} (2013) 81--104}, \href{http://arxiv.org/abs/1208.3025}{{\ttfamily arXiv:1208.3025 [astro-ph.CO]}}.

\bibitem{Peter:2012jh}
A.~H.~G. Peter, M.~Rocha, J.~S. Bullock, and M.~Kaplinghat, ``{Cosmological Simulations with Self-Interacting Dark Matter II: Halo Shapes vs. Observations},'' \href{http://dx.doi.org/10.1093/mnras/sts535}{{\em Mon. Not. Roy. Astron. Soc.} {\bfseries 430} (2013) 105}, \href{http://arxiv.org/abs/1208.3026}{{\ttfamily arXiv:1208.3026 [astro-ph.CO]}}.

\bibitem{Elbert:2014bma}
O.~D. Elbert, J.~S. Bullock, S.~Garrison-Kimmel, M.~Rocha, J.~O\~norbe, and A.~H.~G. Peter, ``{Core formation in dwarf haloes with self-interacting dark matter: no fine-tuning necessary},'' \href{http://dx.doi.org/10.1093/mnras/stv1470}{{\em Mon. Not. Roy. Astron. Soc.} {\bfseries 453} no.~1, (2015) 29--37}, \href{http://arxiv.org/abs/1412.1477}{{\ttfamily arXiv:1412.1477 [astro-ph.GA]}}.

\bibitem{Fry:2015rta}
A.~B. Fry, F.~Governato, A.~Pontzen, T.~Quinn, M.~Tremmel, L.~Anderson, H.~Menon, A.~M. Brooks, and J.~Wadsley, ``{All about baryons: revisiting SIDM predictions at small halo masses},'' \href{http://dx.doi.org/10.1093/mnras/stv1330}{{\em Mon. Not. Roy. Astron. Soc.} {\bfseries 452} no.~2, (2015) 1468--1479}, \href{http://arxiv.org/abs/1501.00497}{{\ttfamily arXiv:1501.00497 [astro-ph.CO]}}.

\bibitem{Kaplinghat:2015aga}
M.~Kaplinghat, S.~Tulin, and H.-B. Yu, ``{Dark Matter Halos as Particle Colliders: Unified Solution to Small-Scale Structure Puzzles from Dwarfs to Clusters},'' \href{http://dx.doi.org/10.1103/PhysRevLett.116.041302}{{\em Phys. Rev. Lett.} {\bfseries 116} no.~4, (2016) 041302}, \href{http://arxiv.org/abs/1508.03339}{{\ttfamily arXiv:1508.03339 [astro-ph.CO]}}.

\bibitem{Tulin:2017ara}
S.~Tulin and H.-B. Yu, ``{Dark Matter Self-interactions and Small Scale Structure},'' \href{http://dx.doi.org/10.1016/j.physrep.2017.11.004}{{\em Phys. Rept.} {\bfseries 730} (2018) 1--57}, \href{http://arxiv.org/abs/1705.02358}{{\ttfamily arXiv:1705.02358 [hep-ph]}}.

\bibitem{Harvey:2015hha}
D.~Harvey, R.~Massey, T.~Kitching, A.~Taylor, and E.~Tittley, ``{The non-gravitational interactions of dark matter in colliding galaxy clusters},'' \href{http://dx.doi.org/10.1126/science.1261381}{{\em Science} {\bfseries 347} (2015) 1462--1465}, \href{http://arxiv.org/abs/1503.07675}{{\ttfamily arXiv:1503.07675 [astro-ph.CO]}}.

\bibitem{Sagunski:2020spe}
L.~Sagunski, S.~Gad-Nasr, B.~Colquhoun, A.~Robertson, and S.~Tulin, ``{Velocity-dependent Self-interacting Dark Matter from Groups and Clusters of Galaxies},'' \href{http://dx.doi.org/10.1088/1475-7516/2021/01/024}{{\em JCAP} {\bfseries 01} (2021) 024}, \href{http://arxiv.org/abs/2006.12515}{{\ttfamily arXiv:2006.12515 [astro-ph.CO]}}.

\bibitem{Harvey:2018uwf}
D.~Harvey, A.~Robertson, R.~Massey, and I.~G. McCarthy, ``{Observable tests of self-interacting dark matter in galaxy clusters: BCG wobbles in a constant density core},'' \href{http://dx.doi.org/10.1093/mnras/stz1816}{{\em Mon. Not. Roy. Astron. Soc.} {\bfseries 488} no.~2, (2019) 1572--1579}, \href{http://arxiv.org/abs/1812.06981}{{\ttfamily arXiv:1812.06981 [astro-ph.CO]}}.

\bibitem{Bondarenko:2017rfu}
K.~Bondarenko, A.~Boyarsky, T.~Bringmann, and A.~Sokolenko, ``{Constraining self-interacting dark matter with scaling laws of observed halo surface densities},'' \href{http://dx.doi.org/10.1088/1475-7516/2018/04/049}{{\em JCAP} {\bfseries 04} (2018) 049}, \href{http://arxiv.org/abs/1712.06602}{{\ttfamily arXiv:1712.06602 [astro-ph.CO]}}.

\bibitem{DES:2023bzs}
{\bfseries DES} {\bfseries Collaboration}, D.~Cross { et~al.}, ``{Examining the self-interaction of dark matter through central cluster galaxy offsets},'' \href{http://dx.doi.org/10.1093/mnras/stae442}{{\em Mon. Not. Roy. Astron. Soc.} {\bfseries 529} no.~1, (2024) 52--58}, \href{http://arxiv.org/abs/2304.10128}{{\ttfamily arXiv:2304.10128 [astro-ph.CO]}}.

\bibitem{Adhikari:2022sbh}
S.~Adhikari { et~al.}, ``{Astrophysical Tests of Dark Matter Self-Interactions},'' \href{http://arxiv.org/abs/2207.10638}{{\ttfamily arXiv:2207.10638 [astro-ph.CO]}}.

\bibitem{Kamada:2017tsq}
A.~Kamada, H.~Kim, and T.~Sekiguchi, ``{Axionlike particle assisted strongly interacting massive particle},'' \href{http://dx.doi.org/10.1103/PhysRevD.96.016007}{{\em Phys. Rev. D} {\bfseries 96} no.~1, (2017) 016007}, \href{http://arxiv.org/abs/1704.04505}{{\ttfamily arXiv:1704.04505 [hep-ph]}}.

\bibitem{Chu:2024rrv}
X.~Chu, M.~Nikolic, and J.~Pradler, ``{Even SIMP miracles are possible},'' \href{http://arxiv.org/abs/2401.12283}{{\ttfamily arXiv:2401.12283 [hep-ph]}}.

\bibitem{Pich:1991fq}
A.~Pich and E.~de~Rafael, ``{Strong CP violation in an effective chiral Lagrangian approach},'' \href{http://dx.doi.org/10.1016/0550-3213(91)90019-T}{{\em Nucl. Phys. B} {\bfseries 367} (1991) 313--333}.

\bibitem{Scherer:2002tk}
S.~Scherer, ``{Introduction to chiral perturbation theory},'' {\em Adv. Nucl. Phys.} {\bfseries 27} (2003) 277, \href{http://arxiv.org/abs/hep-ph/0210398}{{\ttfamily arXiv:hep-ph/0210398}}.

\bibitem{DiLuzio:2020wdo}
L.~Di~Luzio, M.~Giannotti, E.~Nardi, and L.~Visinelli, ``{The landscape of QCD axion models},'' \href{http://dx.doi.org/10.1016/j.physrep.2020.06.002}{{\em Phys. Rept.} {\bfseries 870} (2020) 1--117}, \href{http://arxiv.org/abs/2003.01100}{{\ttfamily arXiv:2003.01100 [hep-ph]}}.

\bibitem{Note1}
Throughout ${\protect \rm Tr}[\lambda _a\lambda _b]=2\delta _{ab}$, while the symmetric tensors are $ d_{abc}=\protect \frac {1}{4}{\protect \rm Tr}(\{\lambda _a,\lambda _b\}\lambda _c)$ , and $c_{abcde}=(\delta _{ab}d_{cde}+\delta _{cd}d_{abe})/N_f+\protect \frac {1}{2}d_{abm}d_{cdn }d_{mne}$.

\bibitem{Kamada:2022zwb}
A.~Kamada, S.~Kobayashi, and T.~Kuwahara, ``{Perturbative unitarity of strongly interacting massive particle models},'' \href{http://dx.doi.org/10.1007/JHEP02(2023)217}{{\em JHEP} {\bfseries 02} (2023) 217}, \href{http://arxiv.org/abs/2210.01393}{{\ttfamily arXiv:2210.01393 [hep-ph]}}.

\bibitem{Planck:2018vyg}
{\bfseries Planck} {\bfseries Collaboration}, N.~Aghanim { et~al.}, ``{Planck 2018 results. VI. Cosmological parameters},'' \href{http://dx.doi.org/10.1051/0004-6361/201833910}{{\em Astron. Astrophys.} {\bfseries 641} (2020) A6}, \href{http://arxiv.org/abs/1807.06209}{{\ttfamily arXiv:1807.06209 [astro-ph.CO]}}. [Erratum: Astron.Astrophys. 652, C4 (2021)].

\bibitem{Kolb:1979qa}
E.~W. Kolb and S.~Wolfram, ``{Baryon Number Generation in the Early Universe},'' \href{http://dx.doi.org/10.1016/0550-3213(82)90012-8}{{\em Nucl. Phys. B} {\bfseries 172} (1980) 224}. [Erratum: Nucl.Phys.B 195, 542 (1982)].

\bibitem{future}
C.~Garc\'\i{}a-Cely, G.~Landini, L.~Marsili, and O.~Zapata {\em In preparation} (2024) .

\bibitem{Frumkin:2021zng}
R.~Frumkin, Y.~Hochberg, E.~Kuflik, and H.~Murayama, ``{Thermal Dark Matter from Freeze-Out of Inverse Decays},'' \href{http://dx.doi.org/10.1103/PhysRevLett.130.121001}{{\em Phys. Rev. Lett.} {\bfseries 130} no.~12, (2023) 121001}, \href{http://arxiv.org/abs/2111.14857}{{\ttfamily arXiv:2111.14857 [hep-ph]}}.

\bibitem{Chu:2018fzy}
X.~Chu, C.~Garcia-Cely, and H.~Murayama, ``{Velocity Dependence from Resonant Self-Interacting Dark Matter},'' \href{http://dx.doi.org/10.1103/PhysRevLett.122.071103}{{\em Phys. Rev. Lett.} {\bfseries 122} no.~7, (2019) 071103}, \href{http://arxiv.org/abs/1810.04709}{{\ttfamily arXiv:1810.04709 [hep-ph]}}.

\bibitem{Chu:2019awd}
X.~Chu, C.~Garcia-Cely, and H.~Murayama, ``{A Practical and Consistent Parametrization of Dark Matter Self-Interactions},'' \href{http://dx.doi.org/10.1088/1475-7516/2020/06/043}{{\em JCAP} {\bfseries 06} (2020) 043}, \href{http://arxiv.org/abs/1908.06067}{{\ttfamily arXiv:1908.06067 [hep-ph]}}.

\bibitem{Note2}
Note that this justifies the expansion in $\theta $ of Eq.~\protect \eqref {eq:chiLagrangian}.

\bibitem{Tran:2024vxy}
V.~Tran, D.~Gilman, M.~Vogelsberger, X.~Shen, S.~O'Neil, and X.~Zhang, ``{Gravothermal Catastrophe in Resonant Self-interacting Dark Matter Models},'' \href{http://arxiv.org/abs/2405.02388}{{\ttfamily arXiv:2405.02388 [astro-ph.GA]}}.

\bibitem{Kamada:2023dse}
A.~Kamada and H.~J. Kim, ``{Evolution of resonant self-interacting dark matter halos},'' \href{http://dx.doi.org/10.1103/PhysRevD.109.063535}{{\em Phys. Rev. D} {\bfseries 109} no.~6, (2024) 063535}, \href{http://arxiv.org/abs/2304.12621}{{\ttfamily arXiv:2304.12621 [astro-ph.CO]}}.

\bibitem{Yang:2022hkm}
D.~Yang and H.-B. Yu, ``{Gravothermal evolution of dark matter halos with differential elastic scattering},'' \href{http://dx.doi.org/10.1088/1475-7516/2022/09/077}{{\em JCAP} {\bfseries 09} (2022) 077}, \href{http://arxiv.org/abs/2205.03392}{{\ttfamily arXiv:2205.03392 [astro-ph.CO]}}.

\bibitem{Yang:2022zkd}
S.~Yang, X.~Du, Z.~C. Zeng, A.~Benson, F.~Jiang, E.~O. Nadler, and A.~H.~G. Peter, ``{Gravothermal Solutions of SIDM Halos: Mapping from Constant to Velocity-dependent Cross Section},'' \href{http://dx.doi.org/10.3847/1538-4357/acbd49}{{\em Astrophys. J.} {\bfseries 946} no.~1, (2023) 47}, \href{http://arxiv.org/abs/2205.02957}{{\ttfamily arXiv:2205.02957 [astro-ph.CO]}}.

\bibitem{Outmezguine:2022bhq}
N.~J. Outmezguine, K.~K. Boddy, S.~Gad-Nasr, M.~Kaplinghat, and L.~Sagunski, ``{Universal gravothermal evolution of isolated self-interacting dark matter halos for velocity-dependent cross-sections},'' \href{http://dx.doi.org/10.1093/mnras/stad1705}{{\em Mon. Not. Roy. Astron. Soc.} {\bfseries 523} no.~3, (2023) 4786--4800}, \href{http://arxiv.org/abs/2204.06568}{{\ttfamily arXiv:2204.06568 [astro-ph.GA]}}.

\bibitem{Hayashi:2020syu}
K.~Hayashi, M.~Ibe, S.~Kobayashi, Y.~Nakayama, and S.~Shirai, ``{Probing dark matter self-interaction with ultrafaint dwarf galaxies},'' \href{http://dx.doi.org/10.1103/PhysRevD.103.023017}{{\em Phys. Rev. D} {\bfseries 103} no.~2, (2021) 023017}, \href{http://arxiv.org/abs/2008.02529}{{\ttfamily arXiv:2008.02529 [astro-ph.CO]}}.

\bibitem{Valli:2017ktb}
M.~Valli and H.-B. Yu, ``{Dark matter self-interactions from the internal dynamics of dwarf spheroidals},'' \href{http://dx.doi.org/10.1038/s41550-018-0560-7}{{\em Nature Astron.} {\bfseries 2} (2018) 907--912}, \href{http://arxiv.org/abs/1711.03502}{{\ttfamily arXiv:1711.03502 [astro-ph.GA]}}.

\bibitem{Correa_2021}
C.~A. Correa, ``Constraining velocity-dependent self-interacting dark matter with the milky way’s dwarf spheroidal galaxies,'' \href{http://dx.doi.org/10.1093/mnras/stab506}{{\em Monthly Notices of the Royal Astronomical Society} {\bfseries 503} no.~1, (Feb., 2021) 920–937}. \url{http://dx.doi.org/10.1093/mnras/stab506}.

\bibitem{Tsai:2020vpi}
Y.-D. Tsai, R.~McGehee, and H.~Murayama, ``{Resonant Self-Interacting Dark Matter from Dark QCD},'' \href{http://dx.doi.org/10.1103/PhysRevLett.128.172001}{{\em Phys. Rev. Lett.} {\bfseries 128} no.~17, (2022) 172001}, \href{http://arxiv.org/abs/2008.08608}{{\ttfamily arXiv:2008.08608 [hep-ph]}}.

\bibitem{Witten:1984rs}
E.~Witten, ``{Cosmic Separation of Phases},'' \href{http://dx.doi.org/10.1103/PhysRevD.30.272}{{\em Phys. Rev. D} {\bfseries 30} (1984) 272--285}.

\bibitem{Schwaller:2015tja}
P.~Schwaller, ``{Gravitational Waves from a Dark Phase Transition},'' \href{http://dx.doi.org/10.1103/PhysRevLett.115.181101}{{\em Phys. Rev. Lett.} {\bfseries 115} no.~18, (2015) 181101}, \href{http://arxiv.org/abs/1504.07263}{{\ttfamily arXiv:1504.07263 [hep-ph]}}.

\bibitem{Helmboldt:2019pan}
A.~J. Helmboldt, J.~Kubo, and S.~van~der Woude, ``{Observational prospects for gravitational waves from hidden or dark chiral phase transitions},'' \href{http://dx.doi.org/10.1103/PhysRevD.100.055025}{{\em Phys. Rev. D} {\bfseries 100} no.~5, (2019) 055025}, \href{http://arxiv.org/abs/1904.07891}{{\ttfamily arXiv:1904.07891 [hep-ph]}}.

\bibitem{Reichert:2021cvs}
M.~Reichert, F.~Sannino, Z.-W. Wang, and C.~Zhang, ``{Dark confinement and chiral phase transitions: gravitational waves vs matter representations},'' \href{http://dx.doi.org/10.1007/JHEP01(2022)003}{{\em JHEP} {\bfseries 01} (2022) 003}, \href{http://arxiv.org/abs/2109.11552}{{\ttfamily arXiv:2109.11552 [hep-ph]}}.

\bibitem{Aoki:2006we}
Y.~Aoki, G.~Endrodi, Z.~Fodor, S.~D. Katz, and K.~K. Szabo, ``{The Order of the quantum chromodynamics transition predicted by the standard model of particle physics},'' \href{http://dx.doi.org/10.1038/nature05120}{{\em Nature} {\bfseries 443} (2006) 675--678}, \href{http://arxiv.org/abs/hep-lat/0611014}{{\ttfamily arXiv:hep-lat/0611014}}.

\bibitem{Bhattacharya:2014ara}
T.~Bhattacharya { et~al.}, ``{QCD Phase Transition with Chiral Quarks and Physical Quark Masses},'' \href{http://dx.doi.org/10.1103/PhysRevLett.113.082001}{{\em Phys. Rev. Lett.} {\bfseries 113} no.~8, (2014) 082001}, \href{http://arxiv.org/abs/1402.5175}{{\ttfamily arXiv:1402.5175 [hep-lat]}}.

\bibitem{Bai:2023cqj}
Y.~Bai, T.-K. Chen, and M.~Korwar, ``{QCD-collapsed domain walls: QCD phase transition and gravitational wave spectroscopy},'' \href{http://dx.doi.org/10.1007/JHEP12(2023)194}{{\em JHEP} {\bfseries 12} (2023) 194}, \href{http://arxiv.org/abs/2306.17160}{{\ttfamily arXiv:2306.17160 [hep-ph]}}.

\bibitem{EPTA:2023fyk}
{\bfseries EPTA, InPTA:} {\bfseries Collaboration}, J.~Antoniadis { et~al.}, ``{The second data release from the European Pulsar Timing Array - III. Search for gravitational wave signals},'' \href{http://dx.doi.org/10.1051/0004-6361/202346844}{{\em Astron. Astrophys.} {\bfseries 678} (2023) A50}, \href{http://arxiv.org/abs/2306.16214}{{\ttfamily arXiv:2306.16214 [astro-ph.HE]}}.

\bibitem{Reardon:2023gzh}
D.~J. Reardon { et~al.}, ``{Search for an Isotropic Gravitational-wave Background with the Parkes Pulsar Timing Array},'' \href{http://dx.doi.org/10.3847/2041-8213/acdd02}{{\em Astrophys. J. Lett.} {\bfseries 951} no.~1, (2023) L6}, \href{http://arxiv.org/abs/2306.16215}{{\ttfamily arXiv:2306.16215 [astro-ph.HE]}}.

\bibitem{NANOGrav:2023gor}
{\bfseries NANOGrav} {\bfseries Collaboration}, G.~Agazie { et~al.}, ``{The NANOGrav 15 yr Data Set: Evidence for a Gravitational-wave Background},'' \href{http://dx.doi.org/10.3847/2041-8213/acdac6}{{\em Astrophys. J. Lett.} {\bfseries 951} no.~1, (2023) L8}, \href{http://arxiv.org/abs/2306.16213}{{\ttfamily arXiv:2306.16213 [astro-ph.HE]}}.

\bibitem{Xu:2023wog}
H.~Xu { et~al.}, ``{Searching for the Nano-Hertz Stochastic Gravitational Wave Background with the Chinese Pulsar Timing Array Data Release I},'' \href{http://dx.doi.org/10.1088/1674-4527/acdfa5}{{\em Res. Astron. Astrophys.} {\bfseries 23} no.~7, (2023) 075024}, \href{http://arxiv.org/abs/2306.16216}{{\ttfamily arXiv:2306.16216 [astro-ph.HE]}}.

\bibitem{NANOGrav:2023hvm}
{\bfseries NANOGrav} {\bfseries Collaboration}, A.~Afzal { et~al.}, ``{The NANOGrav 15 yr Data Set: Search for Signals from New Physics},'' \href{http://dx.doi.org/10.3847/2041-8213/acdc91}{{\em Astrophys. J. Lett.} {\bfseries 951} no.~1, (2023) L11}, \href{http://arxiv.org/abs/2306.16219}{{\ttfamily arXiv:2306.16219 [astro-ph.HE]}}.

\bibitem{Han:2023olf}
C.~Han, K.-P. Xie, J.~M. Yang, and M.~Zhang, ``{Self-interacting dark matter implied by nano-Hertz gravitational waves},'' \href{http://arxiv.org/abs/2306.16966}{{\ttfamily arXiv:2306.16966 [hep-ph]}}.

\bibitem{Bringmann:2023opz}
T.~Bringmann, P.~F. Depta, T.~Konstandin, K.~Schmidt-Hoberg, and C.~Tasillo, ``{Does NANOGrav observe a dark sector phase transition?},'' \href{http://dx.doi.org/10.1088/1475-7516/2023/11/053}{{\em JCAP} {\bfseries 11} (2023) 053}, \href{http://arxiv.org/abs/2306.09411}{{\ttfamily arXiv:2306.09411 [astro-ph.CO]}}.

\bibitem{Witten:1980sp}
E.~Witten, ``{Large N Chiral Dynamics},'' \href{http://dx.doi.org/10.1016/0003-4916(80)90325-5}{{\em Annals Phys.} {\bfseries 128} (1980) 363}.

\bibitem{DiVecchia:1980yfw}
P.~Di~Vecchia and G.~Veneziano, ``{Chiral Dynamics in the Large n Limit},'' \href{http://dx.doi.org/10.1016/0550-3213(80)90370-3}{{\em Nucl. Phys. B} {\bfseries 171} (1980) 253--272}.

\bibitem{Sannino:2016sfx}
F.~Sannino, A.~Strumia, A.~Tesi, and E.~Vigiani, ``{Fundamental partial compositeness},'' \href{http://dx.doi.org/10.1007/JHEP11(2016)029}{{\em JHEP} {\bfseries 11} (2016) 029}, \href{http://arxiv.org/abs/1607.01659}{{\ttfamily arXiv:1607.01659 [hep-ph]}}.

\bibitem{LANDINIGIACOMO2022DMag}
G.~Landini, {\em Dark Matter and gauge dynamics}.
\newblock PhD thesis, Pisa University, 2022.

\bibitem{Buttazzo:2019mvl}
D.~Buttazzo, L.~Di~Luzio, P.~Ghorbani, C.~Gross, G.~Landini, A.~Strumia, D.~Teresi, and J.-W. Wang, ``{Scalar gauge dynamics and Dark Matter},'' \href{http://dx.doi.org/10.1007/JHEP01(2020)130}{{\em JHEP} {\bfseries 01} (2020) 130}, \href{http://arxiv.org/abs/1911.04502}{{\ttfamily arXiv:1911.04502 [hep-ph]}}.

\bibitem{Witten:1982fp}
E.~Witten, ``{An SU(2) Anomaly},'' \href{http://dx.doi.org/10.1016/0370-2693(82)90728-6}{{\em Phys. Lett. B} {\bfseries 117} (1982) 324--328}.

\bibitem{Cline:2017qpe}
J.~M. Cline, K.~Kainulainen, and D.~Tucker-Smith, ``{Electroweak baryogenesis from a dark sector},'' \href{http://dx.doi.org/10.1103/PhysRevD.95.115006}{{\em Phys. Rev. D} {\bfseries 95} no.~11, (2017) 115006}, \href{http://arxiv.org/abs/1702.08909}{{\ttfamily arXiv:1702.08909 [hep-ph]}}.

\bibitem{Carena:2018cjh}
M.~Carena, M.~Quir\'os, and Y.~Zhang, ``{Electroweak Baryogenesis from Dark-Sector CP Violation},'' \href{http://dx.doi.org/10.1103/PhysRevLett.122.201802}{{\em Phys. Rev. Lett.} {\bfseries 122} no.~20, (2019) 201802}, \href{http://arxiv.org/abs/1811.09719}{{\ttfamily arXiv:1811.09719 [hep-ph]}}.

\end{thebibliography}\endgroup

\clearpage

\newpage
\onecolumngrid

\appendix

\section{Chiral Lagrangians in a  $\theta$-vacuum}

In the main text we illustrate the effects of a non-vanishing $\theta$ angle in a dark $SU(N_c)$ QCD-like theory with $N_f$ flavors of light fermions in the fundamental representation and $N_c\geq 3$. Our results can be easily generalized to other choices of the gauge group, namely $SO(N_c)$ or $Sp(N_c)$. We first summarize the $SU(N_c)$ case and then discuss the other possibilities.

\textbf{Unitary groups.}
The Lagrangian describing $N_f$ massless Dirac fermions in the fundamental representation of a  $SU(N_c)$ gauge group with $N_c\geq3$ enjoys a global flavor symmetry, $G_F=SU(N_f)_L\times SU(N_f)_R$, under which the fermion chiral components transform as $q_{L(R)}\to \exp{(i\alpha_{L(R)}^a\lambda^a)}q_{L(R)}$, where $\lambda^a$ are the generators of $SU(N_f)$ normalized as $\tr(\lambda^a\lambda^b)=2\delta^{ab}$. Confinement of gauge interactions induces a fermion condensate $\med{\Bar{q}q}\sim\Lambda^3$ which spontaneously breaks the flavor group $SU(N_f)_L\otimes SU(N_f)_R\to SU(N_f)_V$. The unbroken vectorial subgroup is defined as the set of transformations for which $\alpha^a_L=\alpha^a_R$.
This gives rise to $N_f^2-1$ Goldstone bosons, $\pi^a$, living in the coset space $(SU(N_f)_L\otimes SU(N_f)_R)/ SU(N_f)_V\sim SU(N_f)$. Their interactions are described by the effective Lagrangian 
   $\L_{\rm eff}^{M=0}={f_\pi^2}\tr[\partial_\mu U^\dagger\partial^\mu U]/4$,
where $U=e^{i\Pi/f_\pi}$, with $\Pi=\pi^a\lambda^a$.
A quark mass matrix $M$ explicitly breaks the flavor symmetry group $G_F$. However, if $m_q\ll\Lambda$, $G_F$ is still a good approximate symmetry, in which case, the small breaking due to the mass matrix is parametrized in the effective Lagrangian as
\begin{equation}\label{eq:chiralLagApp}
	\L_{\rm eff}=\frac{f_\pi^2}{4}\tr[\partial_\mu U^\dagger\partial^\mu U]+\frac{f_\pi^2}{2}B_0\tr[M^\dagger_\td U+ U^\dagger  M_\td]\,.
\end{equation}
Here, we include the effect 
of a non-vanishing $\theta$ term as $M_\theta=e^{i\theta Q/2}M e^{i\theta Q/2}$, with $Q=M^{-1}/\tr M^{-1}$. The $\pi^a$ acquire a small mass $m_\pi\sim\sqrt{m_q\Lambda}\ll\Lambda$, becoming pseudo-Goldstone bosons.

The presence of $\theta\neq0$ generates interactions involving an odd number of mesons. The leading terms are
\begin{equation}\label{eq:oddtermsApp2}
\L_{\theta}=-\frac{\td}{6f_\pi\tr M^{-1}}B_0\tr\Pi^3+\frac{\theta}{120f_\pi^3\tr M^{-1}}B_0\tr\Pi^5,
\end{equation}
which gives rise to decays $\pi^a\to\pi^b\pi^c$, as well as 3-to-2 processes $\pi^a\pi^b\pi^c\to\pi^d\pi^e$.  We introduce the $SU(N)$ structure constants: the totally anti-symmetric
$f_{abc}=-i\tr([\lambda_a,\lambda_b]\lambda_c)/4$ and the totally symmetric $d_{abc}=\tr(\{\lambda_a,\lambda_b\}\lambda_c)/4$. Using the $SU(N)$ identity $\tr(\lambda_a\lambda_b\lambda_c)=2(d_{abc}+if_{abc})$ and the symmetry properties of the structure constants we get 
\begin{equation}
\!\!\!\L_{\td}=-\frac{B_0\theta}{3f_\pi\tr M^{-1}}\bigg(d_{abc}\pi_a\pi_b\pi_c-\frac{c_{abcde}}{10f_\pi^2}\pi_a\pi_b\pi_c\pi_d\pi_e\bigg),
 \label{eq:thetatermsApp}
\end{equation}
where $c_{abcde}=(\delta_{ab}d_{cde}+\delta_{cd}d_{abe})/N_f+\frac{1}{2}d_{abm}d_{cdn }d_{mne}$. These interactions are present whenever the $d_{abc}$ tensor is non-vanishing, which occurs if $N_f\geq3$.  

Finally, the Lagrangian must be augmented with the well-known Wess-Zumino-Witten (WZW) term, describing the 5-point interaction
\begin{equation}
{\cal L}_{\rm WZW} = \frac{ N_c}{240 \pi^2 f_\pi^5} \epsilon^{\mu \nu \rho \sigma} \operatorname{Tr}\left[\Pi \partial_\mu \Pi \partial_\nu \Pi \partial_\rho \Pi \partial_\sigma \Pi\right]\,.
    \label{eq:WZ}
\end{equation}
The WZW term induces an extra contribution to 3-to-2 processes $\pi^a\pi^b\pi^c\to\pi^d\pi^e$, which competes with  that of $\theta$ and is non-vanishing if the coset space has a non-trivial fifth homotopy group, $\pi_5$. For unitary gauge groups, this occurs when $N_f\geq3$, as in such a case $\pi_5(SU(N_f))=\mathbb{Z}$.

The case $N_c=2$ is special as the flavor group is $SU(2N_f)$, whose corresponding  fermion condensate breaks to $Sp(2N_f)$, giving rise to $2N_f^2-N_f-1$ pseudo-Goldstone bosons. Similar to the case of  $Sp(N_c)$ discussed below, for $N_f \geq 3$ the $d_{abc}$ tensors do not vanish, while  WZW term  appears already for $N_f\geq 2$.

\textbf{Orthogonal groups.} The Lagrangian of $N_f$ massless Weyl fermions in the fundamental representation of $SO(N_c)$ enjoys a $SU(N_f)$ flavor symmetry. As usual, we consider  $N_c\geq4$ because  $N_c=2$ and $N_c=3$ respectively correspond to $U(1)$ and $SU(2)$. The fermion condensate $\med{qq}$ spontaneously breaks the flavor symmetry as $SU(N_f)\to SO(N_f)$, leading to $N_f(N_f+1)/2-1$ pseudo-Goldstone bosons. Their dynamics is described again by Eq.~\eqref{eq:chiralLagApp}  with the $\lambda^a$ matrices in the $U$ field replaced with the elements of the coset $SU(N_f)/SO(N_f)$, which are the real generators of $SU(N_f)$. Note that the real generators of $SU(N_f)$ are symmetric while the imaginary ones are antisymmetric and correspond to those of $SO(N_f)$.
The odd interactions in Eq.~\eqref{eq:thetatermsApp} are present if the symmetric $d_{abc}$ tensor is non-vanishing, which occurs if $N_f\geq3$. The WZW term is non-vanishing also for $N_f\geq3$ as  in that case $\pi_5 (SU(N_f)/SO(N_f))=\mathbb{Z}$~\cite{Hochberg:2014kqa,Sannino:2016sfx}.

\textbf{Symplectic groups.}  The Lagrangian of $N_f$ massless Weyl fermions in the fundamental representation of  a $Sp(N_c)$ gauge group enjoys a $SU(N_f)$ flavor symmetry, which is broken as $SU(N_f)\to Sp(N_f)$ by the fermion condensate. Notice that $N_c$ needs to be an even integer for symplectic groups, and without loss of generality, $N_c\geq 4$ as $Sp(2)=SU(2)$. Furthermore, $N_f$ must be even to avoid the global anomaly of the gauge group~\cite{Witten:1982fp}. The number of pseudo-Goldstone bosons is $N_f(N_f-1)/2-1$.  The interactions among the mesons are described again by Eq.~\eqref{eq:chiralLagApp}  with the meson field replaced by $U=e^{i\pi^a\lambda^a/f_\pi}\gamma_{N_f}$, where $\gamma_{N_f}\equiv \textbf{1}_{N_f/2}\otimes i\sigma_2$ is the antisymmetric invariant tensor of $Sp(N_f)$.
Here, the $\lambda^a$ matrices are the elements of the coset  $SU(N_f)/Sp(N_f)$ which can be expressed as $\{\Tilde{\lambda}_R^a\otimes\textbf{1}_2,\Tilde{\lambda}_I^a\otimes\sigma_k\}/\sqrt{2}$, where $\Tilde{\lambda}_{R(I)}^a$ are the real symmetric (imaginary anti-symmetric) generators of $SU(N_f/2)$ and $\sigma_k$ are the Pauli matrices, see e.g. Refs.~\cite{LANDINIGIACOMO2022DMag, Buttazzo:2019mvl}. 
The $d_{abc}$ tensor entering the odd interactions is non-vanishing for $N_f\geq6$, see e.g. Ref.~\cite{Kamada:2017tsq}, while the WZW term is non-vanishing for $N_f\geq4$  as $\pi_5 (SU(N_f)/Sp(N_f))=\mathbb{Z}$ in that case~\cite{Hochberg:2014kqa}.

\section{Explicit Lagrangian for each benchmark}

\textbf{Tadpoles.} The interactions described in the chiral Lagrangian of Eq.~\eqref{eq:chiralLagApp} induce tadpoles, i.e. linear terms in the meson field $\Pi=\pi^a\lambda^a$. 
At leading order in $\theta$, these arise in the form $\tr[\Pi\{M,Q\}]$. Choosing $Q=M^{-1}/\tr M^{-1}$, which satisfies $\tr Q=1$,
and using $\tr\Pi=0$, the linear terms vanish and Lagrangian takes the form of Eq.~(3) of the main text up to  ${\cal O}(\td^2)$ corrections. At higher orders in $\theta$, one must choose a more involved function $Q(\theta)$, which reduces to the expression above for $\theta\ll1$. 

\begin{figure}[t]
\includegraphics[width=0.45\textwidth]{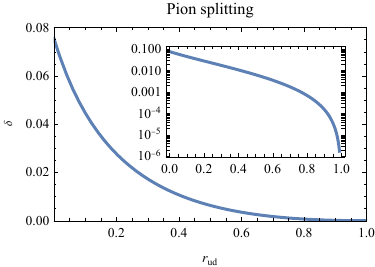}
\includegraphics[width=0.44\textwidth]{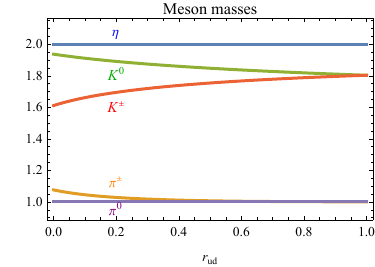}
\caption{{\em Left:} Splitting of the masses of the pions $\delta=m_{\pi^\pm}/m_{\pi^0}-1$  as a function of $r_{ud}$. {\em Right:} Masses of the 8 mesons normalized to $m_{\pi^0}$ as a function of $r_{ud}$.
In both plots $v_R\lesssim0.1$. }
\label{fig:masses}  
 \end{figure}

\textbf{Mass spectrum.} 
The mass spectrum of the mesons is obtained expanding  at the quadratic order in $\Pi$, which gives $\L_{\rm mass}=-B_0\pi^a\pi^b\tr[M\{\lambda^a,\lambda^b\}]/4 + {\cal O } (\theta^2)$. 
In the following we present the details for the two benchmark models discussed in the main text. 
\begin{itemize}
    \item \emph{Benchmark BM1.}
We fix $N_f=3$ and we adopt the same notation of ordinary QCD.
The quark mass matrix is $M={\rm diag}(m_u,m_d,m_s)$ with $m_u\leq m_d\leq m_s$.
The spectrum consists of 8 pseudo-Goldstone bosons: $\pi^\pm=(\pi_1\pm i \pi_2)/\sqrt{2}$, $K^\pm=(\pi_4\pm i \pi_5)/\sqrt{2}$, $K^0/\Bar{K}^0=(\pi_6\pm i\pi_7)/\sqrt{2}$ and~\cite{Scherer:2002tk}
\begin{equation}
\bigg(
    \begin{array}{c}
        \pi^0   \\
        \eta
    \end{array}
    \bigg)
    =
    \bigg(
    \begin{array}{cc}
       \cos{\theta_{\eta\pi}}  &  \sin{\theta_{\eta\pi}}\\
       -\sin{\theta_{\eta\pi}}  & \cos{\theta_{\eta\pi}}
    \end{array}
    \bigg)
    \bigg(
    \begin{array}{c}
        \pi_3   \\
        \pi_8
    \end{array}
    \bigg), \qquad \text{ with } \qquad \tan(2\theta_{\eta\pi})=\frac{\sqrt{3}(m_u-m_d)}{(m_u+m_d-2m_s)}.
\end{equation}

The masses squared of the mesons are $m^2_{\pi^\pm}=B_0(m_u+m_d)$, $m^2_{K^\pm}=B_0(m_u+m_s)$, $m^2_{K,\bar{K}^0}=B_0(m_d+m_s)$,
while $m^2_{\pi^0}$ and $m_\eta^2$ are the eigenvalues of
\begin{equation}
\mathcal{M}^2_{\pi^0,\eta} = \bigg(
    \begin{array}{cc}
       B_0(m_u+m_d)  &  {B_0}(m_u-m_d)/{\sqrt{3}}\\
       {B_0}/(m_u-m_d)/{\sqrt{3}} & {B_0}(m_u+m_d+4m_s)/3
    \end{array}
    \bigg).
\end{equation}
The parameter space of interest in this work is that of resonant scattering,  $m_\eta=(2+v_R^2/4)m_{\pi^0}$ with $v_R\lesssim0.1$ constraining the full spectrum: up to a negligible dependence on $v_R$, the ratios among the meson masses are fixed in term of the parameter $r_{ud}=m_u/m_d$.  In particular, this implies that $m_{\pi^\pm}=(1+\delta)m_{\pi^0}$, with $\delta$ varying from 0 to 0.075, as Fig.~\ref{fig:masses} shows.

\item{\emph{Benchmark BM2.}}
 We fix $N_f=n+1$ with $n\geq3$ and  choose the quark mass matrix as $M ={\rm diag}(m,..m,\mu)$ with $0<m<\mu$. 
There is a remnant $SU(n)$ symmetry under which the $N_f^2-1$ mesons organize as  $n^2-1$ in the adjoint of $SU(n)$ --the pions $\pi$-- as well as $2n$ in the (anti)fundamental representations --the kaons $K$--and one singlet corresponding to the $\eta$ resonance. Their masses squared are respectively $m^2_{\pi}=2B_0 m$, $m^2_{K}= B_0 (m+\mu)$ and $m^2_\eta=2B_0 \left(m+n \mu \right)/(n+1)$. 
The expression $m_\eta=(2+v_R^2/4)m_{\pi}$ translates to 
$ \mu= m[(n+1)(2+v_R^2/4)^2-1]/n$,
which, for $v_R\lesssim0.1$, implies $m_K^2=B_0m(5+3/n)+\mathcal{O}(v_R^2)$. For $N_f=4$ this corresponds to $\mu\simeq5m$, $m_\eta\simeq 2m_\pi$, $m_K\simeq \sqrt{3}m_\pi$.
\end{itemize}

\textbf{Cubic interactions induced by $\theta$.} 
The interactions in Eq.~\eqref{eq:thetatermsApp} predict the decay of the $\eta$ meson. The relevant structure constants are $d_{888}=-d_{338}=-1/\sqrt{3}$ in the BM1 model, while  $d_{\pi_i\pi_i\eta}=\sqrt{2/n(n+1)}$, with $i=1,\dots,n^2-1$, in the BM2 model. The corresponding cubic terms are

\begin{equation}\label{eq:cubicBM1}
\mathcal{L}^{(\rm BM1)}_{\eta\pi\pi}= -\frac{B_0\theta}{\sqrt{3}f_\pi\tr M^{-1}} \cos(3\theta_{\eta\pi})\eta\pi^0\pi^0, \qquad \mathcal{L}^{(\rm BM2)}_{\eta\pi\pi}=- \frac{B_0\theta}{\sqrt{n(n+1)/2} f_\pi\tr M^{-1}} \,
\eta\, \pi \cdot \pi.
\end{equation}
Notice that the expression for BM2 reduces to the one in BM1 if $n=2$ and for vanishing mixing, $\theta_{\eta\pi}=0$.

\section{Boltzmann equations and Dark Matter relic density}

In this section we describe the computation of the DM relic density in more detail.
We focus on the parameter space for which $v_R\lesssim0.1$. 
In this regime, our results are independent on the precise value of $v_R$. 
All the stable mesons contribute to the  DM relic density, although their contribution depends on their mass, as in usual scenarios of co-annihilating DM, where the contribution of the heavier particles is exponentially suppressed.

\textbf{Derivation of the Boltzmann equations.} In general,  the spectrum  can be grouped in the DM particle and the other stable mesons --collectively  referred to as $\DM$-- and the $\eta$ resonance, which decays via $\theta$-induced interactions. As usual, we introduce the parameter $z=m_{\rm DM}/T$ and the yield as $Y_X=n_X/s$, where $n_X$ is the number density and $s=2\pi^2g_sT^3/45$ is the total entropy density of the Universe. The Hubble rate in a radiation-dominated Universe is $H\simeq1.66\sqrt{g_*}T^2/M_{\rm Pl}$ ( $g_*$ being the number of relativistic d.o.f.).
All the stable mesons convert to DM by means of  processes that maintain chemical equilibrium among the different species. This allows us to write a Boltzmann equation for the sum of the yields of DM and its co-annihilating partners, $Y_{\DM}$.
Then, the evolution of $\DM$ and $\eta$ populations is described by the following coupled Boltzmann equations:
\begin{equation}\label{eq:BEQ1}
\!\!\!\!\!\!\!  \begin{cases}
    \begin{split}
        sHz\frac{dY_{\DM}}{dz}=&-\gamma_3({\DM\DM\DM}\to {\DM\DM})\left(\frac{Y_{\DM}^3}{Y_{{\DM},{\rm eq}}^3}-\frac{Y_{\DM}^2}{Y_{{\DM},{\rm eq}}^2}\right)\\
    &\!\!+2\gamma_D(\eta\to {\DM}{\DM})\left(\frac{Y_\eta}{Y_{\eta,{\rm eq}}}-\frac{Y_{\DM}^2}{Y_{{\DM},{\rm eq}}^2}\right)+\gamma_2(\eta\DM\to {\DM}{\DM})\left(\frac{Y_\eta}{Y_{\eta,{\rm eq}}}\frac{Y_{\DM}}{Y_{{\DM},{\rm eq}}}-\frac{Y_{\DM}^2}{Y_{{\DM},{\rm eq}}^2}\right)
    \end{split}
        \\
 \begin{split}
    sHz\frac{dY_\eta}{dz}=&
    &-\gamma_D(\eta\to {\DM}{\DM})\left(\frac{Y_\eta}{Y_{\eta,{\rm eq}}}-\frac{Y_{\DM}^2}{Y_{{\DM},{\rm eq}}^2}\right)-\gamma_2(\eta{\DM}\to {\DM}{\DM})\left(\frac{Y_\eta}{Y_{\eta,{\rm eq}}}\frac{Y_{\DM}}{Y_{{\DM},{\rm eq}}}-\frac{Y_{\DM}^2}{Y_{{\DM},{\rm eq}}^2}\right),\\
        \end{split}
    \end{cases}
\end{equation}
where $\gamma(I)$ is the equilibrium interaction rate density for the process $I$, obtained by adding the individual contributions of the co-annihilating partners.  For $2\to2$ scatterings,  the latter is defined  as
\begin{equation}
    \gamma(12\to34)=S_{12}S_{34}\int d^3p_1d^3p_2f_1^{\rm eq}f_2^{\rm eq}\int d^3p_3d^3p_4(2\pi)^4\delta^4(p_1+p_2-p_3-p_4)|\mathcal{A}_{12\to34}|^2,
\end{equation}
where $S_{12}(S_{34})=1\,(1/2)$ if the initial (final) particles are different (identical), $f^{\rm eq}_i$ is the equilibrium distribution for the particle $i$ and $\mathcal{A}$ is the amplitude of the process.
In the non-relativistic limit, $m_{1,2}\gg T$, this reduces to  
 $   \gamma(12\to 34) {\simeq} S_{12}n_{1,{\rm eq}}n_{2,{\rm eq}}\med{\sigma_{12\to 34} v},
$
in terms of the thermal averaged cross section
$\med{\sigma_{12\to34}v}$ and the equilibrium number densities.
Similar expressions hold for  $n\to2$ processes with $n\geq1$. 
Detailed balance implies $\gamma_2(\DM\DM\to\eta\DM)=\gamma_2(\eta{\DM}\to {\DM}{\DM})$ and similarly for the other processes. Additional number-changing interactions, such as 4-to-2  annihilations are assumed negligible. Note that might not be the case for keV DM~\cite{Bernal:2015xba}.

Furthermore, for $v_R$ sufficiently small, 3-to-2 annihilations  are dominated by the \emph{on-shell} exchange of $\eta$ (see Fig.~1 of the main text). To avoid double-counting, from the total 3-to-2 rate, one must subtract the resonant piece, associated with $\DM\DM \to \eta$ followed by $\eta\DM\to \DM\DM$,  see e.g. Ref.~\cite{Kolb:1979qa}. Here,  $\gamma_3(\DM\DM\DM\to\DM\DM)$ only refers to the non-resonant piece after subtraction, which is small.
In the same region of the parameter space, the inverse decay processes $\DM\DM\to\eta$ are strongly enhanced. As a result decays $\eta\to\DM\DM$ and inverse decay keep the $\DM$ fields and $\eta$ in chemical equilibrium,  even after all other DM number-changing interactions have frozen-out.
In terms of chemical potentials, $\mu_\eta=2\mu_{\DM}$ which leads to
\begin{equation}\label{eq:detailedBalance}
    \frac{Y_\eta}{Y_{\eta,{\rm eq}}}=\frac{Y_{\DM}^2}{Y_{{\DM},{\rm eq}}^2}.
\end{equation}
This only works as long as the  decays and inverse decays $\eta\leftrightarrow\DM\DM$  are both active, which places a lower bound $\theta\gtrsim \theta_{\rm min}$, as discussed in the main text and detailed further below. 
We underline that the presence of the resonance is crucial, as for $m_\eta\gg m_{\rm DM}$, the inverse decay rate gets suppressed by a factor $e^{-m_\eta/m_{\rm DM}}$, freezing out much before than the other DM number-changing processes, invalidating Eq.~\eqref{eq:detailedBalance}. 
All in all, the Boltzmann Eqs.~\eqref{eq:BEQ1}, under the condition Eq.~\eqref{eq:detailedBalance} simplify to
\begin{equation}
    \begin{cases}
    \begin{split}
        sHz\frac{dY_{\DM}}{dz}=& -\bigg(\gamma_2(\eta{\DM}\to {\DM}{\DM})+\gamma_3({\DM\DM\DM}\to {\DM\DM})\bigg)\left(\frac{Y_{\DM}^3}{Y_{{\DM},{\rm eq}}^3}-\frac{Y_{\DM}^2}{Y_{{\DM},{\rm eq}}^2}\right)
        \end{split}
        \\
 \begin{split}
    sHz\frac{dY_\eta}{dz}=&-\gamma_2(\eta{\DM}\to {\DM}{\DM})\left(\frac{Y_{\DM}^3}{Y_{{\DM},{\rm eq}}^2}-\frac{Y_{\DM}^2}{Y_{{\DM},{\rm eq}}^2}\right)\\
        \end{split}
    \end{cases}.
\end{equation}
As expected, the processes $\eta\DM\to\DM\DM$ change the number of DM particles as the 3-to-2 reactions do. 
We can then write a single Boltzmann equation for the combination $ Y_{\DM}+2Y_\eta$,
\begin{equation}
\begin{split}
   sHz\frac{d(Y_{\DM}+2Y_\eta)}{dz}\simeq&-\gamma_2(\eta{\DM}\to {\DM}{\DM})\left(\frac{Y_{\DM}^3}{Y_{{\DM},{\rm eq}}^3}-\frac{Y_{\DM}^2}{Y_{{\DM},{\rm eq}}^2}\right),
\end{split}
\end{equation}
where we neglect the non-resonant piece, $\gamma_3(\DM\DM\DM\to\DM\DM)$. Furthermore, as $Y_{\eta}\ll Y_{{\DM}}$, we can approximate $Y_{\DM}+2Y_\eta\simeq Y_{\DM}$.
In the non-relativistic limit, $ m_{\rm DM}\gg T$, we can write $\gamma_2\simeq s^2Y_{{\DM},{\rm eq}}Y_{\eta,{\rm eq}}\med{\sigma_{\eta\pi} v}$, with $\sigma_{\eta\pi} \equiv\sigma(\eta\DM\to \DM\DM)$, so that
\begin{equation}\label{eq:BEQfinal2}
    \frac{dY_{\DM}}{dz}\simeq-\med{\sigma_{\eta\pi} v}\frac{sY_{\eta,\rm eq}}{zH}\left(\frac{Y_{\DM}^3}{Y_{{\DM},{\rm eq}}^2}-\frac{Y_{\DM}^2}{Y_{{\DM},{\rm eq}}}\right).
\end{equation}
The DM relic density is obtained upon numerical integration of the previous equation. As long as the condition $\theta\gtrsim \theta_{\rm min}$ is full-filled, the DM relic density is independent of the precise value of $\theta$. Our results are well approximated by evaluating the DM relic density as $Y_{\DM,\rm eq}(z_{\rm fo})$ where $z_{\rm fo}$  is the time at which the scatterings $\eta\DM\to\DM\DM$  freeze out, estimated as $H(z_{\rm fo})\approx n_{\eta,{\rm eq}}(z_{\rm fo})\med{\sigma_{\eta\pi}v}$.
This gives (assuming constant $g_s\simeq g_*$)
\begin{equation}
    z_{\rm fo}\simeq \frac{m_{\rm DM}}{m_\eta}\log\left[\frac{\sqrt{z_{\rm fo}}}{1.66\sqrt{g_*}}\left(\frac{m_\eta}{2\pi m_{\rm DM}}\right)^{3/2}m_{\rm DM} \mpl \med{\sigma_{\eta\pi} v}\right],
\end{equation}
which can be solved iteratively. We find $z_{\rm fo}\sim 15-25$.

\textbf{Effective cross sections.}  The annihilation cross section $\med{\sigma_{\eta\pi}}$ depends on the specific benchmark model.
 \begin{itemize}
\item \emph{Benchmark BM2.} The total DM relic density is the sum of the contributions of pions $\pi$ and kaons $K$, which are maintained in chemical equilibrium through  2-to-2 scatterings, which freeze out after $z_{\rm fo}$. Nonetheless,  kaons are heavier and their abundance is exponentially suppressed with respect to the one of the pions as $Y_K/Y_{\pi}\sim \exp{[-(m_K-m_{\rm DM})z/m_{\rm DM}]}$. In the region of the parameter space that we are considering here, $m_K\simeq \sqrt{3}m_\pi$ and the exponential suppression is strong (roughly a factor $\lesssim 10^{-3}$ for $z>10$). Kaons can thus be safely neglected and $Y_{\DM}\simeq Y_\pi$. The only cross section relevant to determine the DM relic density is $\sigma(\eta\pi\to\pi\pi)\equiv\sigma_{\eta\pi}$. As a result, $\med{\sigma_{\eta\pi}v}=\sqrt{5} m_{\rm DM}^2(n^2-4)/(192\pi f_\pi^4 n^2(n+1))$.
\item \emph{Benchmark BM1.}
Here the spectrum is  more involved: four kaons and $\pi^\pm$, which are stable, in addition to the lightest $\pi^0$ and the $\eta$ resonance. All the stable mesons are maintained in chemical equilibrium by conversion processes such as $\pi^+\pi^-\leftrightarrow\pi^0\pi^0$ (and similar for the other species), which freeze-out after $z_{\rm fo}$. While the contribution of kaons is irrelevant in analogy to the BM2 model,  $\pi^\pm$ cannot be neglected, as $\delta\equiv m_{\pi^\pm}/m_{\pi^0}-1\lesssim 0.075$. In fact, at the freeze-out  of $\eta\DM\to\DM\DM$, $\pi^\pm$ are almost as abundant as $\pi^0$. Therefore, $Y_{\DM}\simeq Y_{\pi^0}+Y_{\pi_1}+Y_{\pi_2}=Y_{\pi^0}+2Y_{\pi^+}$. We must take into account the $\eta\pi\to\pi\pi$ processes which involve all the 3 pions. At the leading order in the $\delta$ parameter these are given by
\begin{align}
&\med{\sigma(\eta\pi^0\to \pi_{1,2}\pi_{1,2}) v}=\frac{529\sqrt{5}\,m_{\pi^0}^2}{5184\pi f_\pi^4}\delta,
&\med{\sigma(\eta\pi_{1,2}\to \pi^{0}\pi_{1,2}) v}=\frac{49\sqrt{5}\, \,m_{\pi^0}^2}{2592\pi f_\pi^4}\delta,\quad
&\med{\sigma(\eta\pi^0\to \pi^{0}\pi^{0})v}=\frac{\sqrt{5} \,m_{\pi^0}^2}{64\pi f_\pi^4}\delta,
\end{align}
which, after accounting for combinatoric factors, are summed as
\begin{equation}
\med{\sigma_{\eta\pi} v}\simeq \frac{1}{3}\bigg(\med{\sigma(\eta\pi^0\to \pi^{0}\pi^{0}) v}+2\med{\sigma(\eta\pi^0\to \pi_{1}\pi_{1})v}+2\med{\sigma(\eta\pi_{1}\to \pi^{0}\pi_{1}) v}\bigg).
\end{equation}
In the non-relativistic limit, this enters $\gamma(\eta \DM \to \DM \DM) = n_{\eta,\rm eq}n_{\DM,\rm eq}\med{\sigma_{\eta\pi}v}$, to be replaced in Eq.~\eqref{eq:BEQfinal2}. 

Notice that today DM is made only by $\pi^0$. The conversion processes $\pi^+\pi^-\to\pi^0\pi^0$ can efficiently deplete the $\pi^\pm$ population as long as the mass splitting is $\delta\gtrsim 10^{-5}$~\cite{Katz:2020ywn}, which is realized in our setup as long as $r_{ud}\lesssim0.95$.
\end{itemize}

\textbf{Determination of $\theta_{\rm min}$} \begin{figure*}[t]
\includegraphics[width=0.518\textwidth]{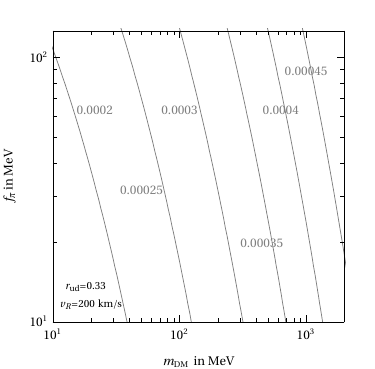}
\includegraphics[width=0.471\textwidth]{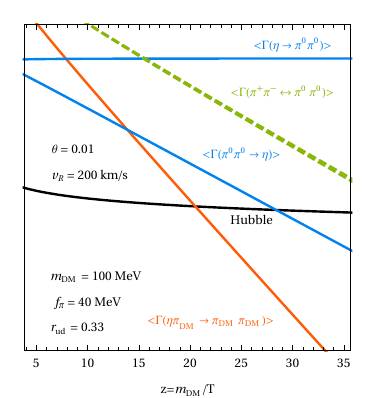}
\caption{{\em Left:}  Contours of  $\theta_{\rm min}$ as a function of $m_{\rm DM}$ and $f_\pi$ in the BM1 model. {\em Right:} Thermal averaged interaction rates for the different processes compared to the Hubble rate, for values of the parameters that reproduce the DM relic abundance with $\theta>\theta_{\rm min}$. Notice that (inverse) decays keep $\eta$ and $\pi^0$ in chemical equilibrium even for $z>z_{\rm fo}$. Analogously $\pi^+\pi^-\leftrightarrow\pi^0\pi^0$ conversions keep the three pions in chemical equilibrium.	} 
 \label{fig:app}
\end{figure*} Eq.~\eqref{eq:detailedBalance} works as long as $\DM$ and $\eta$ are kept in chemical equilibrium after the freeze-out of the reactions $\eta\DM\to\DM\DM$, occurring at $z_{\rm fo}$. 
This requires that $\theta$ is larger than a minimal value, which can be estimated as follows.
We consider the thermal averaged decay and inverse decay rates%
\begin{equation}
\med{\Gamma(\eta\to\DM\DM)}=\frac{K_1(m_\eta/T)}{K_2(m_\eta/T)}\Gamma(\eta\to\DM\DM), \qquad \med{\Gamma(\DM\DM\to\eta)}=\frac{n_{\eta,{\rm eq}}}{n_{{\DM}, {\rm eq}}}\med{\Gamma(\eta\to\DM\DM)},
\end{equation}
where $K_{1,2}$ are the modified Bessel functions. These processes are active  as long as
\begin{equation}
\med{\Gamma(\eta\to\DM\DM)}> H \qquad \text{and} \qquad \med{\Gamma(\DM\DM\to\eta)}>H.
\end{equation}
Since inverse decay rate is exponentially suppressed after $z\gtrsim1$, to maintain the chemical equilibrium among $\eta$ and $\DM$ at $z\geq z_{\rm fo}$, it is necessary that
\begin{equation}
\med{\Gamma(\DM\DM\to\eta)}\geq H \qquad \text{at} \qquad z=z_{\rm fo}.
\end{equation}
This determines the minimal value $\theta_{\rm min}$  as a function of $(v_R,m_{\rm DM},f_\pi)$. We show some reference values in Fig.~\ref{fig:app} (left) for the BM1 model. Similar results hold  the BM2 model.
Notice that the dependence on $v_R$ is not trivial: in the relevant region of the parameter space for this work, $v_R\lesssim 0.1$, we have $\Gamma(\DM\DM\to\eta)\propto v_R$. Hence, as $v_R$ gets smaller, larger values of $\theta_{\rm min}$ are required to compensate the suppression. 
In Fig.~\ref{fig:app} 
(right) we show the interaction rates for the different relevant processes compared with the Hubble rate for different values of the parameters in the BM1 model.
Clearly, the DM production mechanism studied here works even for small values of the $\theta$ parameter, in the range of $10^{-4}$.

\end{document}